\documentclass[manuscript]{aastex}

\shorttitle{Near-IR Imaging Polarimetry toward Bright-Rimmed Clouds}
\shortauthors{Kusune et al.}

\begin{document}

\title{Near-IR Imaging Polarimetry toward a Bright-Rimmed Cloud: Magnetic Field in SFO 74}

\author{Takayoshi Kusune\altaffilmark{1}, Koji Sugitani\altaffilmark{1}, Jingqi Miao\altaffilmark{2}, 
Motohide Tamura\altaffilmark{3,4}, Yaeko Sato\altaffilmark{4}, Jungmi Kwon\altaffilmark{3,4}, Makoto Watanabe\altaffilmark{5},
Shogo Nishiyama\altaffilmark{6}, Takahiro Nagayama\altaffilmark{7}, and Shuji Sato\altaffilmark{8}
}

\altaffiltext{1}{Graduate School of Natural Sciences, Nagoya City University, Mizuho-ku, Nagoya 467-8501, Japan}
\altaffiltext{2}{Centre for Astrophysics \& Planetary Science, School of Physical Sciences, University of Kent, Canterbury, Kent CT2 7NR, UK}
\altaffiltext{3}{Department of Astronomy, The University of Tokyo, 7-3-1 Hongo, Bunkyo-ku, Tokyo 113-0033, Japan}
\altaffiltext{4}{National Astronomical Observatory, 2-21-1 Osawa, Mikata, Tokyo 181-8588, Japan}
\altaffiltext{5}{Department of Cosmosciences, Hokkaido University, Kita 10, Nishi 8, Kita-ku, Sapporo, Hokkaido 060-0810, Japan}
\altaffiltext{6}{Faculty of Education, Miyagi University of Education, Sendai 980-0845, Japan}
\altaffiltext{7}{Department of Physics, Kagoshima University, 1-21-35 Korimoto, Kagoshima 890-0065, Japan}
\altaffiltext{8}{Department of Astrophysics, Nagoya University, Chikusa-ku, Nagoya 464-8602, Japan}

\begin{abstract}
We have made near-infrared ($JHK_{\rm s}$) imaging polarimetry of a bright-rimmed cloud (SFO~74). 
The polarization vector maps clearly show that 
the magnetic field in the layer just behind the bright rim is running along the rim, 
quite different from its ambient magnetic field. 
The direction of the magnetic field just behind the tip rim is almost perpendicular to that of the incident UV radiation, 
and the magnetic field configuration appears to be symmetric as a whole with respect to the cloud symmetry axis. 
We estimated the column and number densities in the two regions (just inside and far inside the tip rim), 
and then derived the magnetic field strength, 
applying the Chandrasekhar-Fermi method. 
The estimated magnetic field strength just inside the tip rim, 
$\sim$90 $\mu$G, 
is stronger than that far inside, 
$\sim$30 $\mu$G. 
This suggests that the magnetic field strength just inside the tip rim is 
enhanced by the UV radiation induced shock. 
The shock increases the density within the top layer 
around the tip, 
and thus increases the strength of the magnetic field. 
The magnetic pressure seems to be comparable to the turbulent one just inside the tip rim, 
implying a significant contribution of the magnetic field to the total internal pressure. 
The mass-to-flux ratio was estimated to be close to the critical value just inside the tip rim.  
We speculate that 
the flat-topped bright rim of SFO~74 could be formed by the magnetic field effect. 
\end{abstract}


\section{INTRODUCTION}
Bright-rimmed clouds (BRCs), 
which are located at the periphery of HII regions, 
are considered to be potential sites for induced star formation 
by UV radiation from nearby OB stars \citep[e.g.,][and references therein]{elme98,elme11}. 
To understand dynamical evolution of such molecular clouds, 
many theorists have developed 2/3-D hydrodynamical models.
They reproduced successfully some of the observed properties of BRCs. 
However, 
most simulations did not always include all of the physical effects related, 
particularly magnetic field effect \citep[e.g.,][]{lef94,white99,will01,kess03,miao09,miao10,mac10}. 
\citet{miao06} did include the magnetic field effect in their 3-D SPH simulation 
by adding a magnetic field pressure in the equation of state. 
However, 
evolution of magnetic field direction cannot be inferred from their simulation. 
In fact, comparing the ambipolar diffusion timescale ($\sim$10 Myr) with the Radiation Driven Implosion (RDI) shock formation ($\sim$ 0.05-0.08 Myr) and 
a BRC formation time (0.1-0.5 Myr), 
it is not difficult to see that magnetic field effect may not be able to produce significant influence generally to the dynamics of the BRC 
because of the extremely low concentration of the ions/electrons in the main body of the BRC. 
Magnetic field may play a certain level of role in morphology formation of a BRC. 
Recently, 
\cite{mac11} did 3-D magnetohydrodynamic simulations and suggested that 
only a strong magnetic field could significantly alter the non-magnetized dynamics.  
More recently, 
\cite{moto13} modeled a photo-evaporating cloud, 
including the magnetic field effect, 
and suggested that the magnetic pressure might play an important role in the cloud density structure. 
It is obvious that the physical effects, 
particularly magnetic field, 
are not treated properly in simulations. 
Thus, 
it is of crucial importance to obtain information on magnetic fields in and around BRCs. 

Wide-field near-infrared (IR) polarimetry is a good tool for measuring the magnetic field direction 
toward molecular clouds \citep[e.g.,][]{tamu07}, 
and \citet{sugi07} made near-IR imaging polarimetry of molecular pillars with bright rims in M~16. 
Their polarization observations show that magnetic 
fields are aligned along some of the pillars in M~16,  
probably due to the UV impact from its exciting stars,
while the global magnetic structure around M~16 is quite different from this alignment.
However, 
such measurements for magnetic field toward BRCs are still rare. 

\begin{figure}
\epsscale{0.41}
\plotone{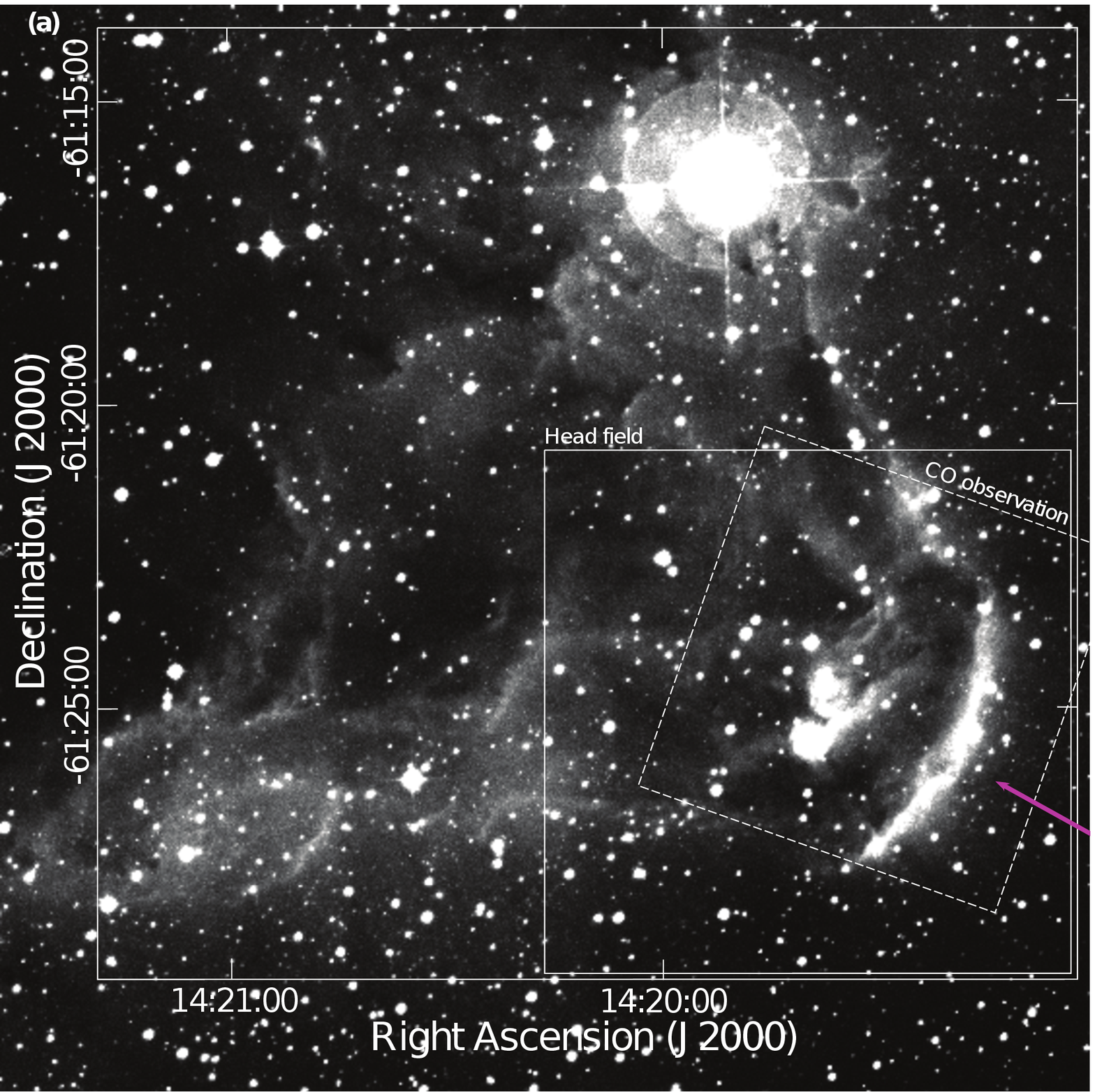}

\plotone{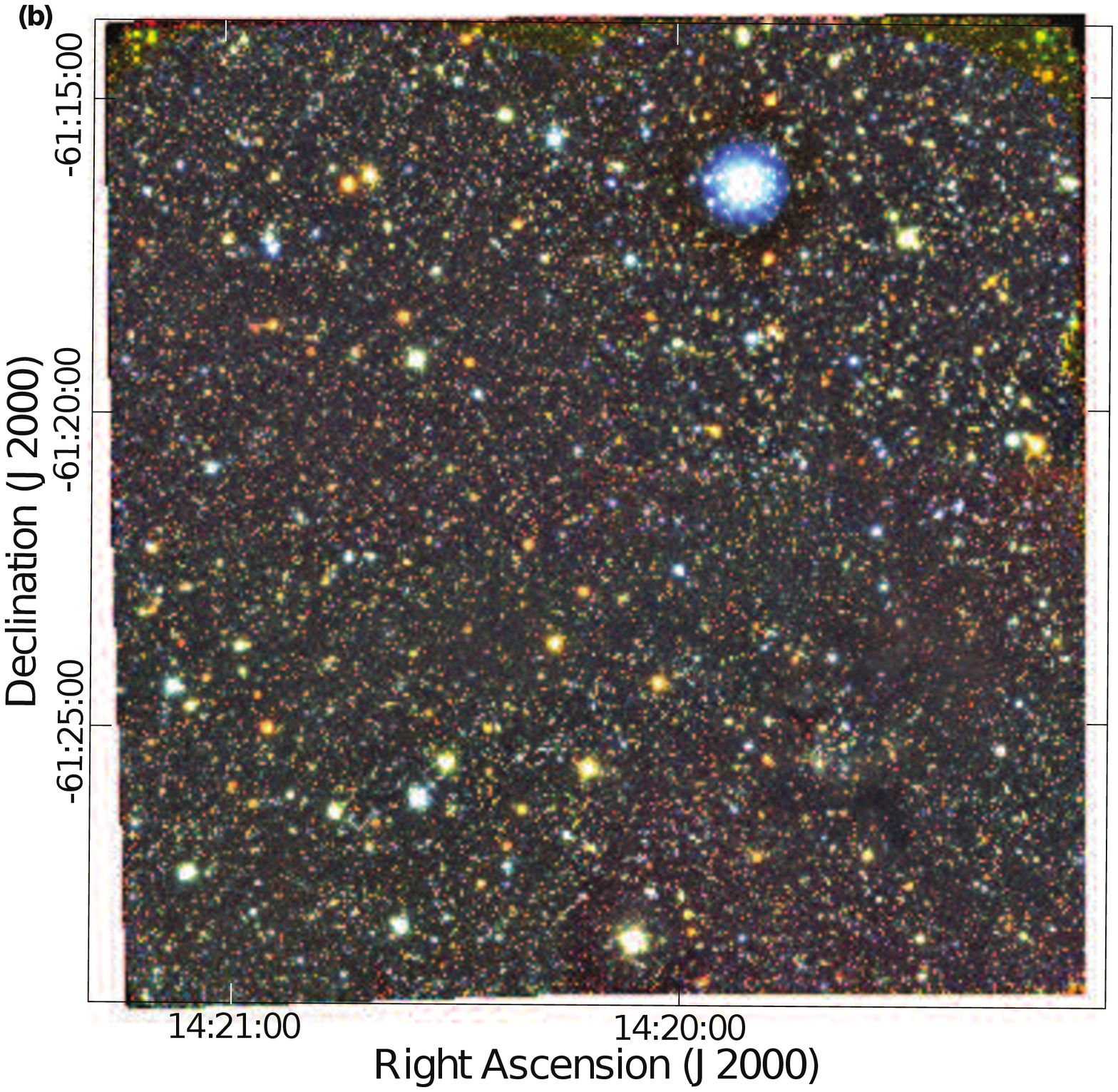}
\caption{
(a) SuperCOSMOS H$_\alpha$ image of SFO~74. 
The incident UV radiation direction from HD~124314, 
which is identified as the main exciting star of RCW~85/SFO~74 
\citep[O6 V;][]{yamar99}, $\sim$60$\degr$ in P.A. is shown with a purple arrow. 
The area of our near-IR imaging polarimetry (see Figure 1(b)) is shown by a solid square with coordinates. 
The head field (each square in Figures \ref{HK-pol}, and \ref{fig9}-\ref{fig11}) is also shown by a small, solid square. 
The area of CO observation is also shown by a dashed square. 
(b) 
Three-color composite, 
mosaic (2$\times$2) image of SFO~74 ($J$, blue; $H$, green; $K_{\rm s}$, red).
The size of the area is $\sim16\farcm0\times16\farcm0$.
The yellow color regions at the upper edge are due to the dead pixels of the $J$-band array.
}
\label{fig1}
\end{figure}

Eighty-nine BRCs associated with IRAS point sources were cataloged by \citet{sugi91} and \citet{sugi94} 
in the northern and southern hemispheres, 
and are referred to as the SFO objects in SIMBAD. 
The levels of star formation of 45 BRCs in the southern hemisphere were examined 
by \citet{urqu09} with CO and mid-infrared data 
as well as the radio continuum data of \citet{thom04}. 
They identified 24 out of them that show strong interaction with HII regions. 
We started 
a systematic study of these BRCs, 
and have so far examined the magnetic fields toward 8 BRCs 
by making a single 
field observation ($\sim$7$\farcm7\times7\farcm7$) of near-IR polarimetry for each BRC. 
In the course of this study, 
we found that the magnetic field of SFO~74 was most clearly revealed among these 8 BRCs. 
Thus, 
in this paper we report the results by adding three more fields around its head field 
(see Figure \ref{fig1}).

SFO~74 is situated in the HII region RCW~85 \citep[$d\sim$1.5 kpc;][]{george87}. 
HD~124314 is identified as the main exciting star of RCW~85/SFO~74 \citep[O6 V;][]{yamar99}, 
with location at the projected distance of $\sim$15 pc from SFO~74. 
We estimated the relatively weak UV photon flux of $\sim$4.5$\times10^{8}$ cm$^{-2}$ s$^{-1}$, 
assuming $\sim$15 pc apart from O6 V star \citep{pana73}, 
nearly the same as that by \citet{urqu09}.
In the head of SFO~74, 
IRAS~14159-6111 point source (MSX~G313.2850-00.3350) is located, 
where \citet{thom04} reported presence of an ultracompact HII (UCHII) region. 
This BRC was classified morphologically as Type A \citep[moderately curved rim;][]{sugi94} 
by focusing on just its brighter tip part. 
As seen in the SuperCOSMOS H$_\alpha$ image \citep{par05} of Figure \ref{fig1}a, 
however, 
its entire shape seems to be Type B (tightly curved rim) rather than Type A.  
This shape appears to be symmetric with respect to the UV incident direction from the exciting stars, 
so it is very likely that this shape was formed by UV radiation, 
i.e., 
the RDI effect by massive stars \citep[e.g.,][]{miao09}. 

In this paper, 
we concentrate on the results of our near-IR polarization measurements of SFO~74, 
including those of the CO archival data. 
In Section 2, 
we describe our near-IR observations and CO archival data. 
In Section 3, 
we present the results of the polarimetry and CO data. 
In Section 4, 
we discuss 
1) the cloud structure and the magnetic field configuration, 
2) the physical quantities related to the magnetic field, 
3) the flat-topped shape of SFO~74 
and 4) star formation activity. 
A summary is given in Section 5. 
Our polarimetric study of other BRCs will be presented in the near future.

\section{OBSERVATIONS AND ARCHIVAL DATA}

\subsection{Near Infrared Observations} 
$JHK_{\rm s}$ polarimetric observations of SFO~74 were made on 2012 February 25 for the head field (a small solid square in Figure \ref{fig1}a). 
Three more fields (north, northeast, and east of the head field) were added on 2013 March 6, 10, 12 and 2014 April 20. 
The mosaic image obtained is shown in Figure \ref{fig1}b. 
The consistency of the polarization angles and degrees were checked among data taken for these two years runs in 2013 and 2014 (see Appendix A). 

We used the imaging polarimeter SIRPOL \citep[polarimetry mode of the SIRIUS camera:][]{kan06} 
mounted on the IRSF 1.4 m telescope at the South African Astronomical Observatory. 
The SIRIUS camera is equipped with three $1024\times1024$ HgCdTe (HAWAII) arrays, $JHK_{\rm s}$ filters, 
and dichroic mirrors, 
which enable simultaneous $JHK_{\rm s}$ observations \citep{nagashima99,nagayama03}. 
The field of view at each band is $\sim$7$\farcm7\times7\farcm7$ with a pixel scale of 0\farcs45. 

We obtained 10 dithered exposures, each 15 s long, at four wave-plate angles 
(0$\degr$, 22$\fdg$5, 45$\degr$ and 67$\fdg$5 in the instrumental coordinate system) 
as one set of observations and repeated this six times. 
Eventually, 
the total on-target exposure times were 900 s per each wave-plate angle. 
Self-sky images were used for median sky subtraction. 
The average seeing size ranged from $\sim$1$\farcs2$ to 1$\farcs7$ at $K_{\rm s}$ during the observations. 
Twilight flat-field images were obtained at the beginning and/or end of the observations. 

\subsection{CO Archival Data}
The $^{12}$CO, $^{13}$CO, and C$^{18}$O  (1--0) data of the Mopra telescope 
were taken from the Australia Telescope Online Archive (Project Codes: M126, Source Name: SFO74) 
in order to reveal the structure of the cloud and to evaluate the velocity dispersions of molecular gas. 
The data are available only within a square area of $\sim$6$\farcm0\times6\farcm0$ toward the cloud head (the dashed square in Figure \ref{fig1}a) and 
was reduced using Gridzilla/Livedata\footnote{http://www.atnf.csiro.au/computing/software/livedata/}. 
According to the web page (Information for Mopra observers\footnote{http://www.narrabri.atnf.csiro.au/mopra/obsinfo.html}), 
the beamwidth (FWMH) is $\sim$30$\arcsec$, 
the antenna efficiency is $\sim$0.4, 
and the velocity resolution is $\sim$0.088 km s$^{-1}$ at 115 GHz .

\section{RESULTS}

\subsection{Photometry and Source Selection}
Standard reduction procedures were applied with IRAF. 
Aperture polarimetry was performed at $J$, $H$ and $K_{\rm s}$ with an aperture radius of 
about FWHM size by using APHOT of the DAOPHOT package. 
The Two Micron All Sky Survey (2MASS) catalog \citep{skrut06} was used for absolute photometric calibration. 
We calculated the Stokes parameters as follows: 
$Q=I_{0}-I_{45}$, $U=I_{22.5}-I_{67.5}$, 
and $I=(I_{0}+I_{22.5}+I_{45}+I_{67.5})/2$, 
where $I_{0}$, $I_{22.5}$, $I_{45}$ and $I_{67.5}$ are intensities at four wave-plate angles. 
To obtain the Stokes parameters in the equatorial coordinate system ($Q'$ and $U'$), 
rotation of $105\degr$ \citep[][and see also Appendix B]{kan06} was applied to them. 
We calculated the degree of polarization $P$ and the polarization angle $\theta$ as follows: 
$P=\sqrt{Q^2+U^2}/I$, and $\theta=(1/2){\rm{tan}}^{-1}(U'/Q')$. 
The absolute accuracy of the position angle (P.A.) of polarization was estimated to be better than $3\degr$ \citep{kan06}, 
where P.A. is the angle measured from north in an easterly direction.
The polarization efficiencies were estimated to be 95.5\%, 96.3\% and 98.5\% at $J$, $H$ and $K_{\rm s}$, respectively, 
and the measurable polarization is $\sim$0.3\% all over the field of view at each band \citep{kan06}. 
With these high polarization efficiencies and low instrument polarization, 
no particular corrections were made here. 
The polarization measurement error ($\Delta P$) and the position angle error ($\Delta\theta$) 
were calculated from the photometric errors, 
and the degrees of polarization were debiased as $P_{\rm debias}=\sqrt{P^2-\Delta P^2}$ \citep{war74}. 
Hereafter, 
we use $P$ as substitute for this debiased value. 
In consideration of the above measurable polarization of $\sim$0.3\%, 
for the sources with the polarization measurement errors of $<0.3\%$ 
we adopt 0.3\% as their polarization errors 
when we calculate $\Delta\theta$. 
We confirmed an agreement of the polarization angle within their errors between two independent observations for almost all the stars 
that were located commonly in the neighboring two fields.

\begin{figure}
\epsscale{0.7}
\plotone{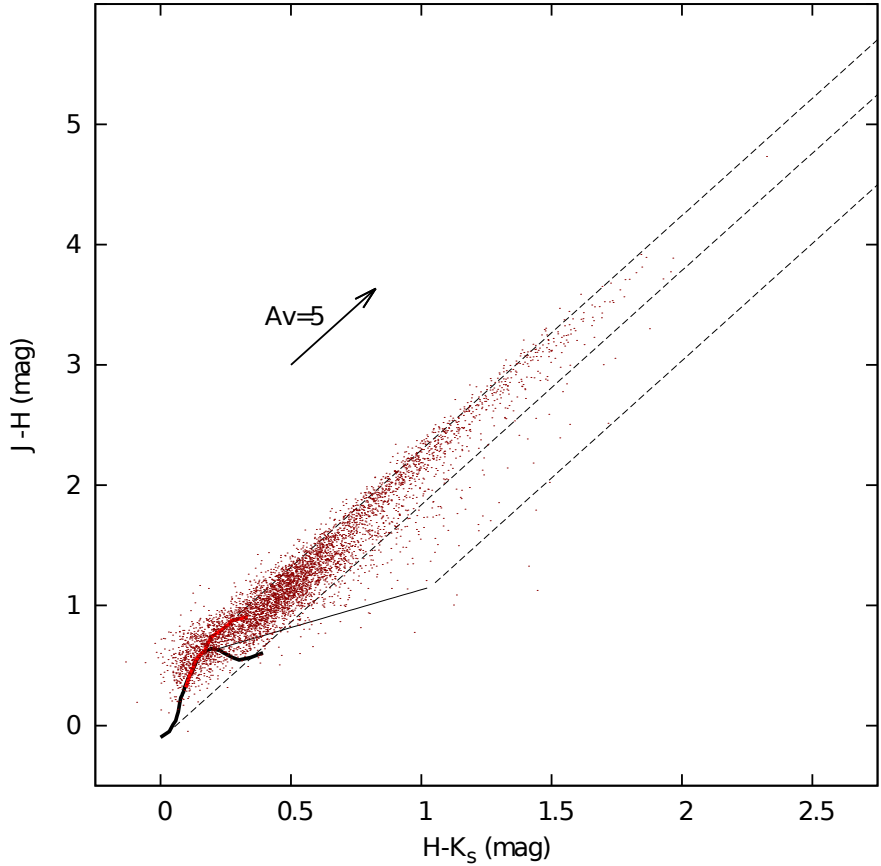}
\caption{
$J-H$ vs. $H-K_{\rm s}$ color-color diagram toward SFO~74. 
The thick curves are the loci of dwarfs (black line) and giants (red line). 
The data for A0-M6 dwarfs and G0-M7 giants are from \citet{bess88}. 
The unreddened CTTS locus of \cite{mey97} is also denoted by the thin solid line. 
These data are transformed to the 2MASS system by using the equations on the explanatory supplement (http://www.ipac.caltech.edu/2mass/releases/allsky/doc/sec6\_4b.html by R. M. Cutri et al.). 
The three dashed lines are reddening ones parallel to the reddening vector \citep[$E(J-H)/E(H-K_{\rm s})=$1.95;][]{chini98}. 
}
\label{fig2}
\end{figure}

We have measured $J$, $H$ and $K_{\rm s}$ polarization for point sources to examine the magnetic field structure.
Only the sources with photometric errors of $<0.1$ mag were used for analysis. 
We excluded sources plotted on red-ward of the reddening line from A0 star on the $J-H$ versus $H-K_{\rm s}$ diagram (Figure \ref{fig2}), 
and/or those associated with reflection nebulae as young stellar object candidates. 
As seen in this color-color diagram, 
most of the detected sources seem to be giants or reddened giants. 
Here, 
we adopted the reddening line of $E({\rm J}-{\rm H})/E({\rm H}-{\rm K}) \sim$ 1.95 \citep{chini98}, 
because this value seems to fit to the distribution of the reddened sources in Figure \ref{fig2}.

\begin{figure}
\epsscale{0.5}
\plotone{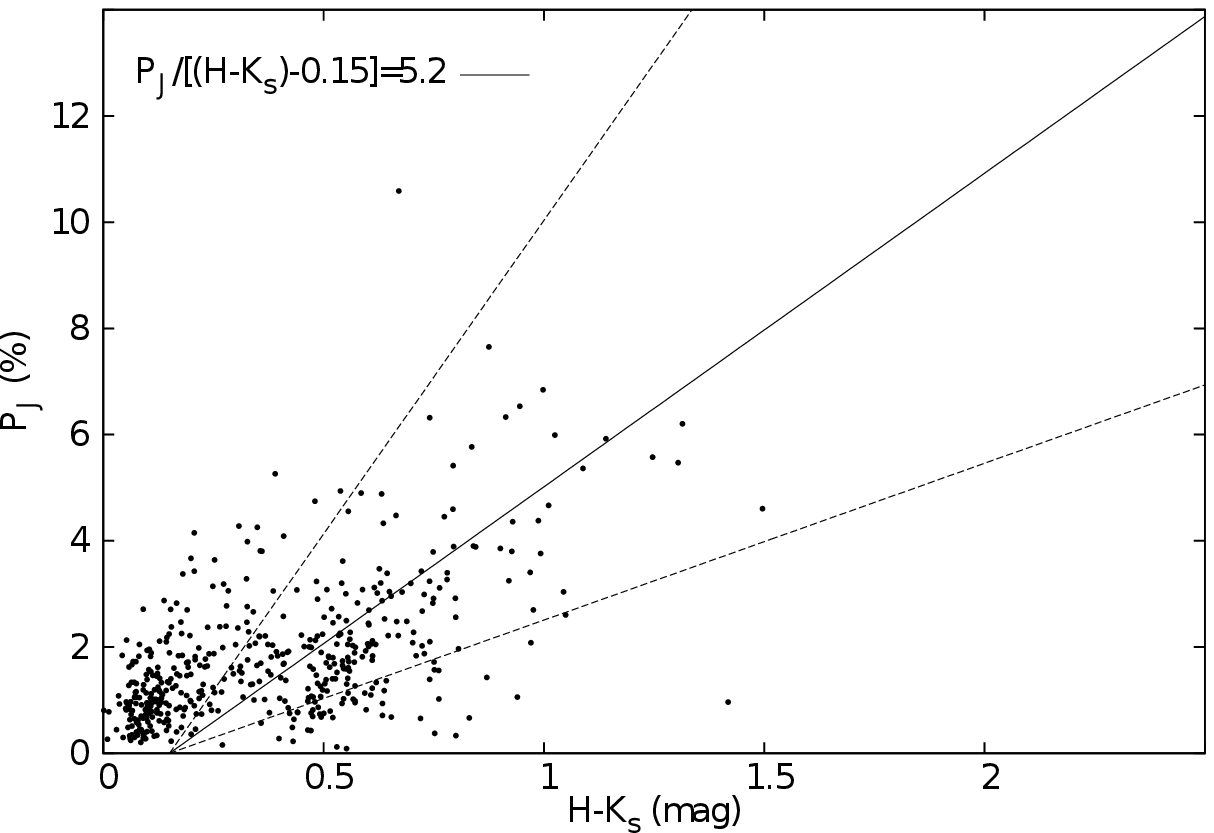}

\plotone{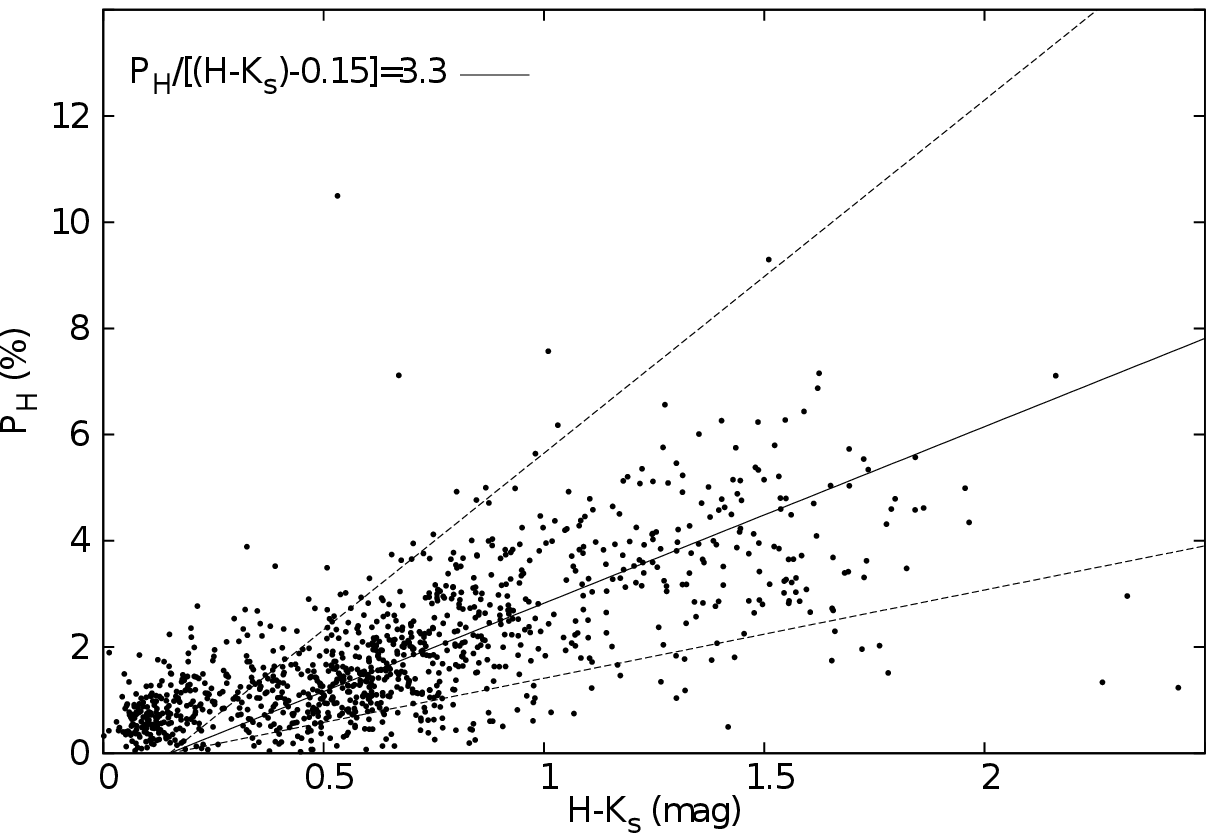}

\plotone{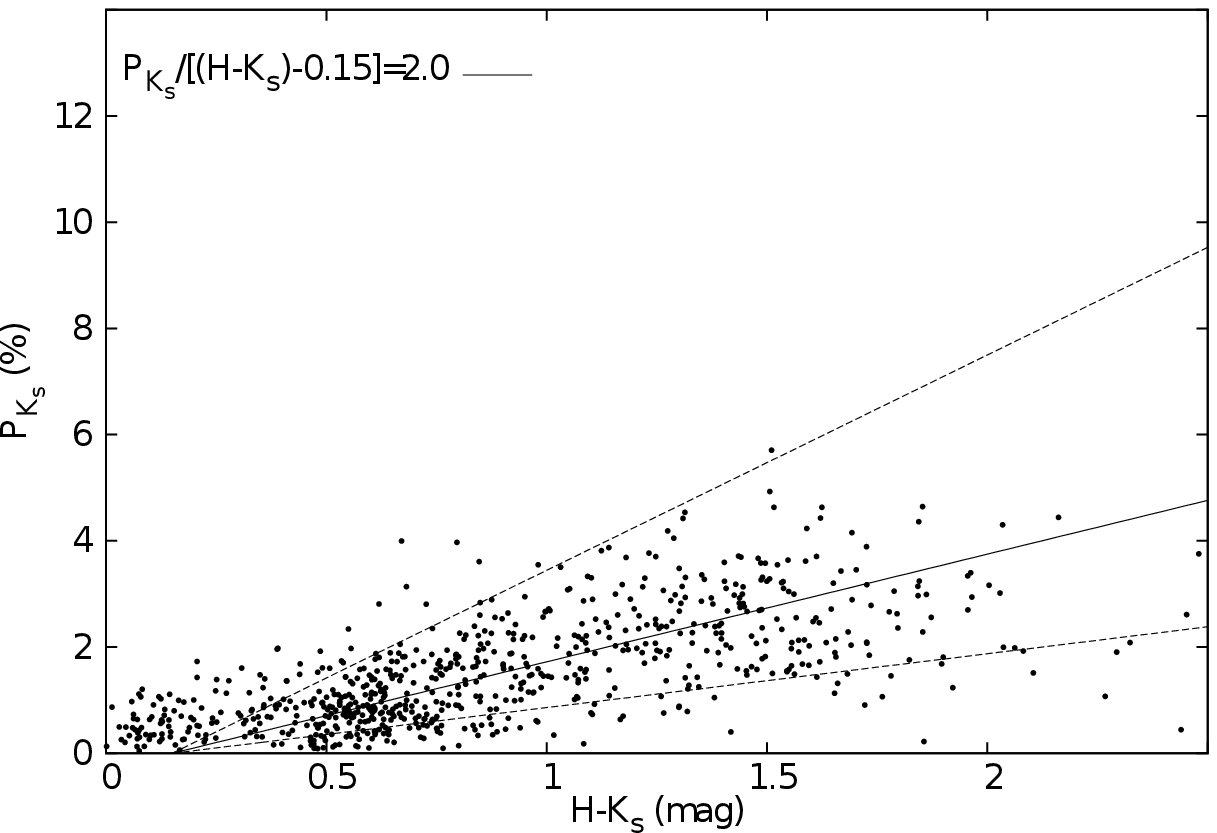}
\caption{
Polarization degree 
vs. $H-K_{\rm s}$ color diagrams at $J$, $H$ and $K_{\rm s}$ for sources having the polarization measurement error of $<0.3\%$. 
The best fit linear lines obtained for the sources with $H-K_{\rm s}>0.15$ are shown by solid lines  at each panels. 
The dashed lines are double and half of the best fit lines, 
and are the lines of our adopted maximum or minimum polarization efficiency. 
}
\label{fig3}
\end{figure}

We regard the polarization of these point sources as the polarization of the dichroic origin, 
and assume the polarization vector to represent the direction of the local magnetic field 
averaged over the line of sight to the source. 
We made polarization degree $P$ versus $H-K_{\rm s}$ color diagrams for point sources with the polarization measurement errors of $<0.3\%$ (Figure \ref{fig3}). 
For the $J$-band panel we include only the sources detected at all three bands, but  
for the $H$- and $K_{\rm s}$-band panels the sources detected both at $H$ and $K_{\rm s}$ bands.
We obtained the best fit linear lines 
$a_J=P(J)/[(H-K_{\rm s})-0.15]$, 
$a_H=P(H)/[(H-K_{\rm s})-0.15]$ and $a_{K_{\rm s}}=P(K_{\rm s})/[(H-K_{\rm s})-0.15]$ (see Figure \ref{fig3}) 
assuming the intrinsic color $(H-K{\rm s})_0$ $=$0.15 of background sources  
based on the model calculations by \citet{wain92}.

As \citet{kusa08} mentioned, 
all of the observed near-IR polarizations are not originated by interstellar polarization, 
i.e., there are highly polarized sources that cannot be explained by interstellar origin, 
but by their circumstellar structures such as disks and envelopes. 
In order to exclude the point sources with large polarization degrees that are not considered to be of dichroic origin, 
we set an upper limit for the interstellar polarization. 
On the other hand,
if the source has intrinsic polarization or background polarization that had occurred  in the more distant area,
and if its polarization direction is different from that of the local magnetic field of interest, 
its polarization degree might become small. 
We also set a lower limit to exclude such sources. 
In Figure \ref{fig3}, 
the slopes of the dashed lines of each panel are double and one-half of the slope of the best fit linear line, and  
most of the plotted sources lie between these two dashed lines. 
Therefore, 
we adopted them as the upper and lower limits of the polarization efficiency. 
Only the sources between the two limits and of $P/\Delta P>3.0$, 
corresponding to $\Delta\theta\la10\degr$, 
were used for plotting polarization vectors. 

We can see a clump of stars located blueward of each adopted upper limit line in Figure \ref{fig3}, 
e.g., 
the one located at $E(H-K_{\rm s})\lesssim0.2$ on the $H$-band panel. 
Most stars of the clump are visible in the optical image,
so they are likely to be foreground dwarfs, 
and contain no information on the magnetic field of SFO~74. 
Thus, 
the clump indicates  
small foreground polarization at $H$, 
so we safely excluded these clumps from the sample. 

\subsection{Polarization Data}
\begin{figure}
\epsscale{0.8}
\plotone{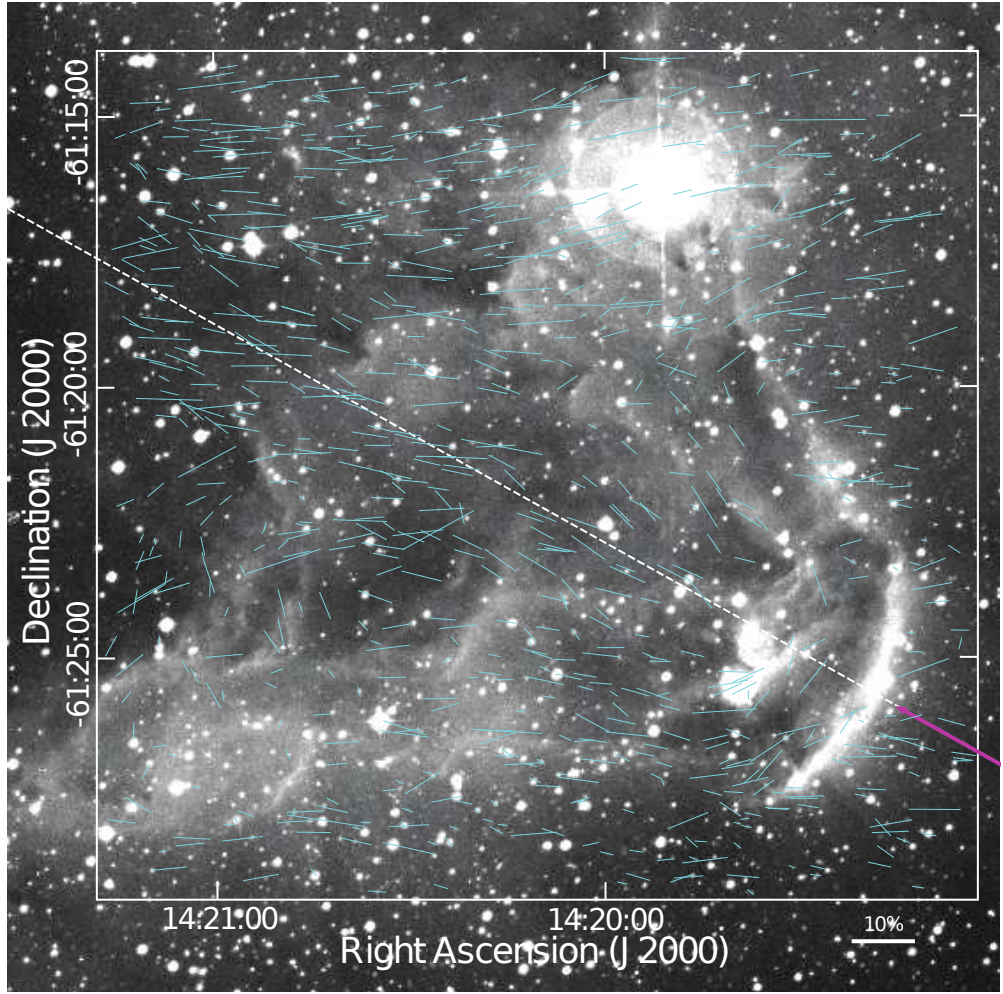}
\caption{
$J$-band polarization vector map of SFO~74 superposed on the SuperCOSMOS H$_\alpha$ image.
A 10\% vector is shown at the right bottom. 
The incident UV radiation direction is shown with a purple arrow, 
and its extended line (our adopted approximately symmetry axis) is shown by a white dashed line. 
}
\label{J-pol}
\end{figure}

Figure \ref{J-pol} shows a $J$-band polarization vector map superposed on the SuperCOSMOS H$\alpha$ image of SFO~74.
SFO~74 is clearly visible in optical, 
so polarization as well as extinction is mostly caused by the cloud or background, 
not foreground interstellar medium. 

\begin{figure}
\epsscale{0.5}
\plotone{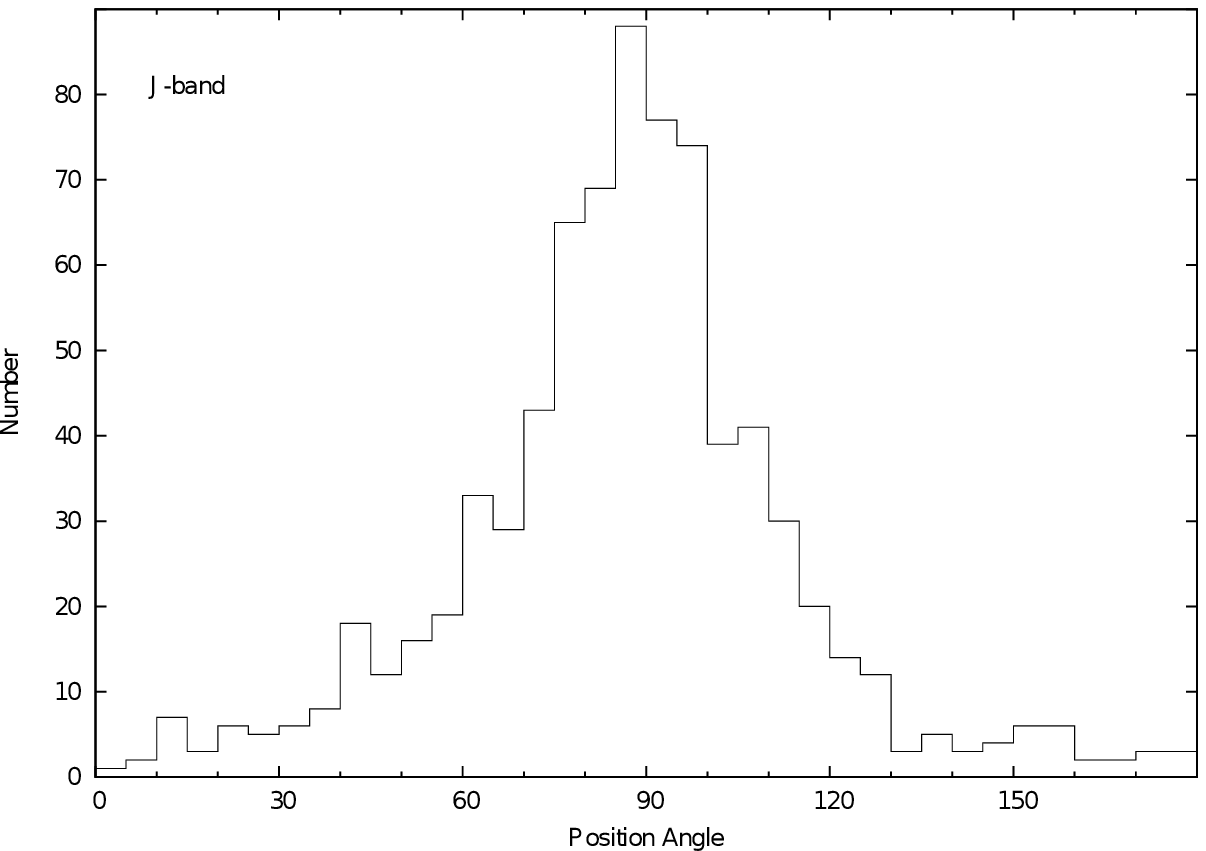}

\plotone{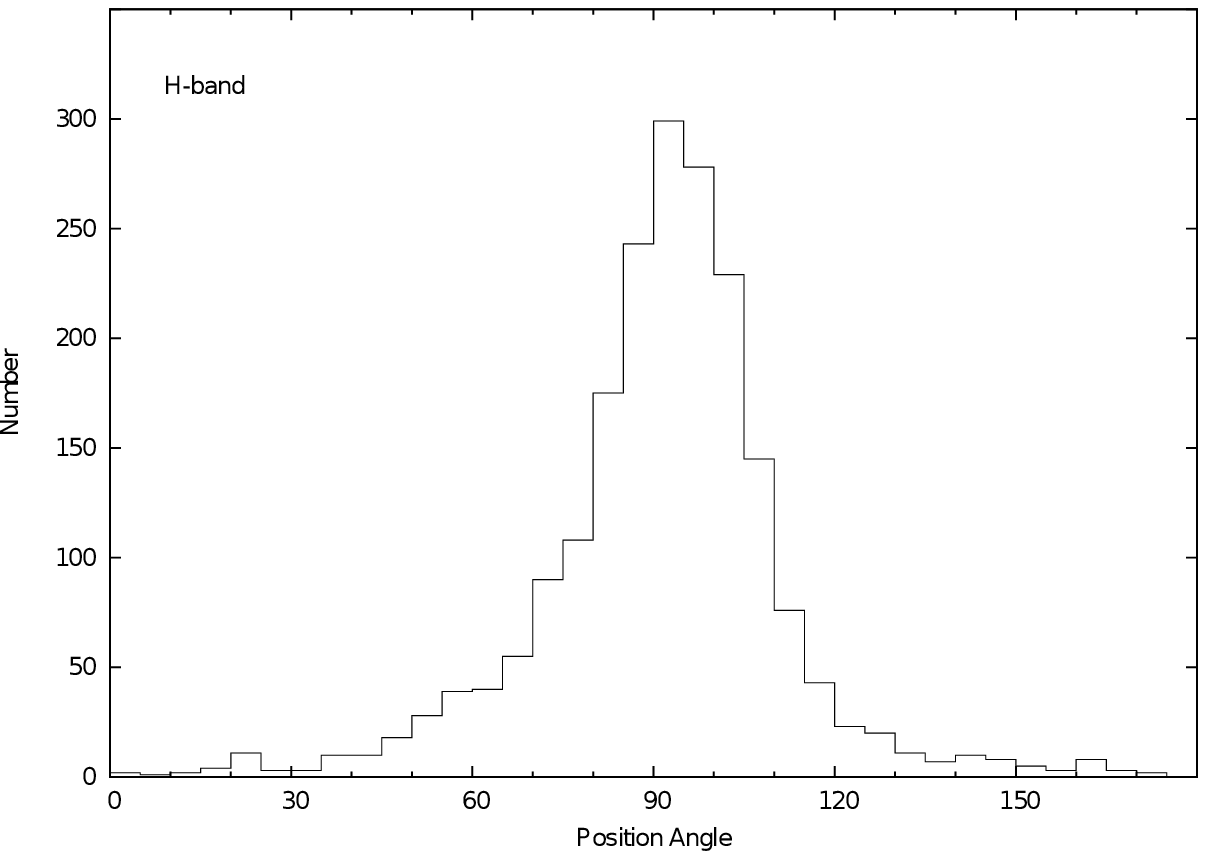}

\plotone{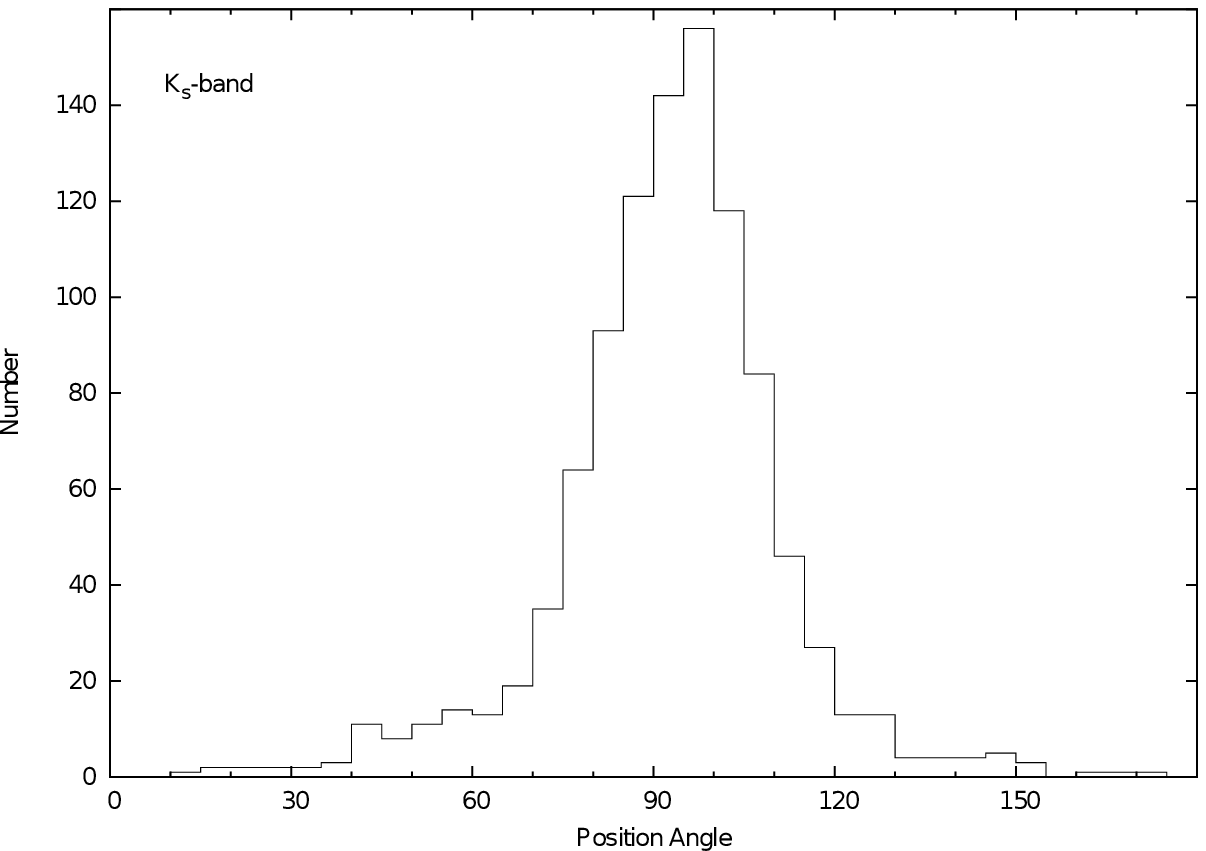}
\caption{
Histograms for the polarization position angles for point sources of $P/\Delta P >3.0$. 
Only the sources lie between our adapted maximum and minimum limits of the polarization efficiencies at $J$-, $H$- and $K_{\rm s}$-band (see Section 3.1). 
For J-band, 
all the sources with polarization vectors in Figure 4 are used here. 
}
\label{fig5}
\end{figure}

In Figure \ref{J-pol}, 
the magnetic field directions inside the cloud seem different from that outside of the cloud (toward the northern, western and southern edges).  
We constructed the histograms of P.A. for $J$-, $H$- and $K_{\rm s}$-band polarization vectors over the entire observed area (Figure \ref{fig5}). 
These histograms have a peak of $\sim$95$\degr$ in P.A., 
indicating the direction of the ambient magnetic field. 
The longer the observed wavelength is, 
the clearer this peak becomes. 
This is because that the extinction, 
i.e., 
the polarization degree, 
decreases at longer wavelengths. 
Then, 
most of the polarization angles at $H$ and $K_{\rm s}$ are likely indicating the direction of the background magnetic field. 

Inside the cloud, 
the polarization vectors follow two patterns; 
the polarization vectors 
1) following the curved bright rim just behind the bright rim and 
2) being parallel to an approximate symmetry axis of the cloud in the area away from the tip bright rim. 
As the symmetry axis, 
we adopt an extension of the line (purple solid line in Figure \ref{J-pol}) connecting the position of HD~124314 and the middle point of the tip bright rim of the cloud (dashed line in Figure \ref{J-pol}). 
These patterns of the magnetic field configuration are even clearer in Figure \ref{HK-pol}, 
showing $H$- and $K_{\rm s}$-band polarization vector maps toward the head region of SFO~74. 

Figure \ref{fig7} shows the change of P.A.s of $H$-band polarization along the symmetric axis. 
We used the sources within the white dashed box in Figure \ref{fig7}a, 
and 
the horizontal axis of Figure \ref{fig7}b represents a position ($x$) from the tip bright rim, 
where the position direction is the tail side of the cloud. 
In a range of $x\sim-2$-$0\arcmin$, an average of P.A. is $\sim$93$\degr$, 
which is identical to the direction of the ambient magnetic field. 
The P.A.s in the tip region just behind the bright rim ($x\sim$ 0-2$\arcmin$) seem to scatter around 140$\degr$, 
and their average is $\sim$141\degr, 
nearly perpendicular to the incident direction of UV radiation, $\sim$60$\degr$. 
In a range of $x\sim$ 2-4$\arcmin$, 
P.A.s decline drastically from $\sim$141$\degr$ to $\sim$70$\degr$.
In $x\sim$ 4-6$\arcmin$, 
an average of P.A. is $\sim$68$\degr$, 
being nearly parallel to the symmetric axis with some scatter.
In $x\sim$ 6-9$\arcmin$, 
P.A.s gradually increase from $\sim$70$\degr$ to $\sim$90$\degr$ and keep constant around $\sim$90$\degr$ thereafter.
In $x \gtrsim$9$\arcmin$, 
most of P.A.s seem to show around $\sim$90$\degr$, 
being close to that of the ambient magnetic field, 
although there is certain number of sources with P.A. $\sim$60$\degr$. 
Excluding the sources with P.A. $<$70$\degr$, 
an average of P.A. is calculated to be $\sim$86$\degr$ in $x\sim$ 9-14$\arcmin$, 
being different only by $\sim$9$\degr$ relative to that of the ambient magnetic field. 
Thus, 
this region with $x\gtrsim9\arcmin$ seems to be dominated by the background polarization component, 
while the region with $x\sim$ 0-6$\arcmin$ seems to be dominated by the cloud polarization one.  
The intermediate region with $x\sim$ 6-9$\arcmin$ can be considered as the transition region.

\subsection{CO Data}
The maps of the  $^{12}$CO peak antenna temperature ($T_{\rm A}^*$),  
the $^{13}$CO integrated intensity, and 
the $^{13}$CO velocity width are shown in Figure 
\ref{fig9}\footnote{Toward this area, two strong velocity components of $\sim$-30 km s$^{-1}$ and $\sim$-50 km s$^{-1}$, 
and other minor components are detectable. 
Here, 
we focus on only the $\sim$-30 km s$^{-1}$ component of SFO~74.}. 
The peak $T_{\rm A}^*$ was determined by Gaussian fitting of $^{12}$CO spectra 
for the points 
detected above 5$\sigma$ level
and the maximum value is $\sim$28 K on the peak $T_{\rm A}^*$ map. 
We derived the excitation temperature $T_{\rm ex}$, and 
the $^{13}$CO column density $N_{^{13}\rm CO}$, 
following the equations of \citet{yaman99} with the antenna efficiency of 0.4. 
We calculated the column density $N_{\rm H_2}$, 
adopting the column density ratio of H$_2$ to $^{13}$CO as $\sim5.0\times10^5$ \citep{dick78b}. 
The peak column density derived is $\sim5.1\times10^{22}$ cm$^{-2}$ at the peak of the $^{13}$CO integrated intensity 
and the column density distribution is very similar to the integrated intensity distribution in morphology. 
Hence, 
we do not show the column density map here.  
The total mass toward the observed was estimated to be $\sim$1400 $M_{\sun}$, 
which is a bit larger than that of $\sim$1000 $M_{\sun}$ estimated by \citet{yamar99}. 
The $^{13}$CO velocity width was determined by Gaussian fitting of $^{13}$CO spectra 
for the points 
detected above 5$\sigma$ level.

In CO, 
the SFO~74 cloud is seen inside the bright rim, 
and its edge on the exciting star side well agrees with the tip bright rim. 
In the $T^*_{\rm A}$ map, 
the temperature is the highest just behind the tip bright rim, 
and gradually decreases with distance away from the tip bright rim. 
The derived maximum $T_{\rm ex}$ is $\sim$70 K, 
which suggests that the cloud temperature is much higher than that in typical dark clouds. 
The peak positions are slightly different between two maps (Figure \ref{fig9}a and \ref{fig9}b) with a separation of $\sim$1$\arcmin$, and 
the peak in the $T^*_{\rm A}$ map is closer to the tip bright rim. 
The IRAS source is not located toward the peak of the $T^*_{\rm A}$ map or that of the $^{13}$CO integrated intensity map. 
In Figure \ref{fig9}c, 
the velocity width is small 
in the area that elongated along/just behind the tip bright rim with an almost uniform value of $\sim$1.4 km s$^{-1}$.
The velocity width increases with distance away from the tip bright rim, 
but eventually decreases. 
The areas of the largest velocity width are located $\sim$2$\arcmin$ from the tip bright rim,  
and do not coincide with the two peaks of the $^{12}$CO $T^*_{\rm A}$ and $^{13}$CO integrated intensity maps.

Figure \ref{fig10} shows the channel maps of $^{13}$CO. 
The LSR central velocity ranges from $-31.17$ km s$^{-1}$ up to $-26.13$ km s$^{-1}$, 
with a velocity interval of 0.46 km s$^{-1}$. 
With increase in velocity, 
the 'mushroom'-like structure (a core + 'earlike' structure at the two sides of the core) becomes visible, 
most clearly at 
$-28.88$ km s$^{-1}$. 
At the channels over $-27.96$ km s$^{-1}$, 
another component appears $\sim$2$\arcmin$ east-southeast of the peak position seen at $-28.88$ km s$^{-1}$. 
In Figure \ref{fig11}a, 
we present the two velocity components superposed on the Spitzer image 
by showing two integrated intensity maps with different velocity ranges. 
Hereafter, 
we refer to the blue component of $\sim-29$ km s$^{-1}$ as 'Cloud~A' 
and to the red component of $\sim-27.5$ km s$^{-1}$ as 'Cloud~B'. 
The IRAS source is located toward Cloud~B, and 
a bright mid-IR nebula of the Spitzer image is also seen toward Cloud~B.
We believe that Cloud~B is located at the near side of Cloud~A 
because a dark area that may correspond to Cloud~B is seen in the optical image (Figure \ref{fig11}b).

\section{ANALYSIS AND DISCUSSION} 

\subsection{Cloud~A Structure and Magnetic Field Configuration} 

The $^{13}$CO channel maps (Figure \ref{fig10}), 
show the 'mushroom' structure consisting of the core and the 'ears'. 
The part of the ears has higher redshifted velocity components with respect to the main velocity component of SFO~74
(see the two panels of $-30.71$ km s$^{-1}$ and $-30.25$ km s$^{-1}$). 
This structure appears to be consistent with the illustration derived from the 3-D simulation results of RDI \citep[Figure 5 of][]{miao09}, 
indicating that the SFO~74 structure had been formed by RDI. 
As already mentioned above, 
the temperature appears to decrease with distance away from the tip bright rim, and 
no sign of star formation is seen  at the peak position of the $T^*_{\rm A}$ map (see the details in Section 4.4). 
These represent that 
Cloud~A is externally heated by the exciting stars of RCW~85, 
and support the idea that RDI process has been going on. 

The magnetic field configuration must have been also affected by RDI. 
In the very early stage of RDI, the magnetic field lines of Cloud~A might be the same as that of the ambient magnetic field. 
When the top layer gas is ionized by the UV radiation of the exciting stars of RCW~85, 
a strong shock is induced by the ionizing/heating and propagates into the cloud. 
Then the surface layer of the cloud gets strongly compressed, 
which causes a magnetic field line form an open loop as shown in Figure \ref{MFC}. 
The magnetic field line of the loop looks almost perpendicular to the UV radiation direction. 
The looped magnetic field lines in the condensed layer form a stream of fragment lines around the bright rim as seen from Figure \ref{HK-pol}. 
In Section 4.3, 
we proof the magnetic pressure could further enhance this field line looping effect. 
With the further propagation into the cloud, 
although the shock becomes weaker, 
it continues driving the gas to converge toward the symmetric axis and eventually to form a dense core-tail structure of type B BRC. 
Farther from the bright rim, 
the shock effect decreases its effect on gas compression (magnetic enhancement). 
We conclude that the magnetic field configuration of SFO~74 is a result of RDI. 

\subsection{Physical Quantities}
\begin{figure}
\epsscale{0.8}
\plotone{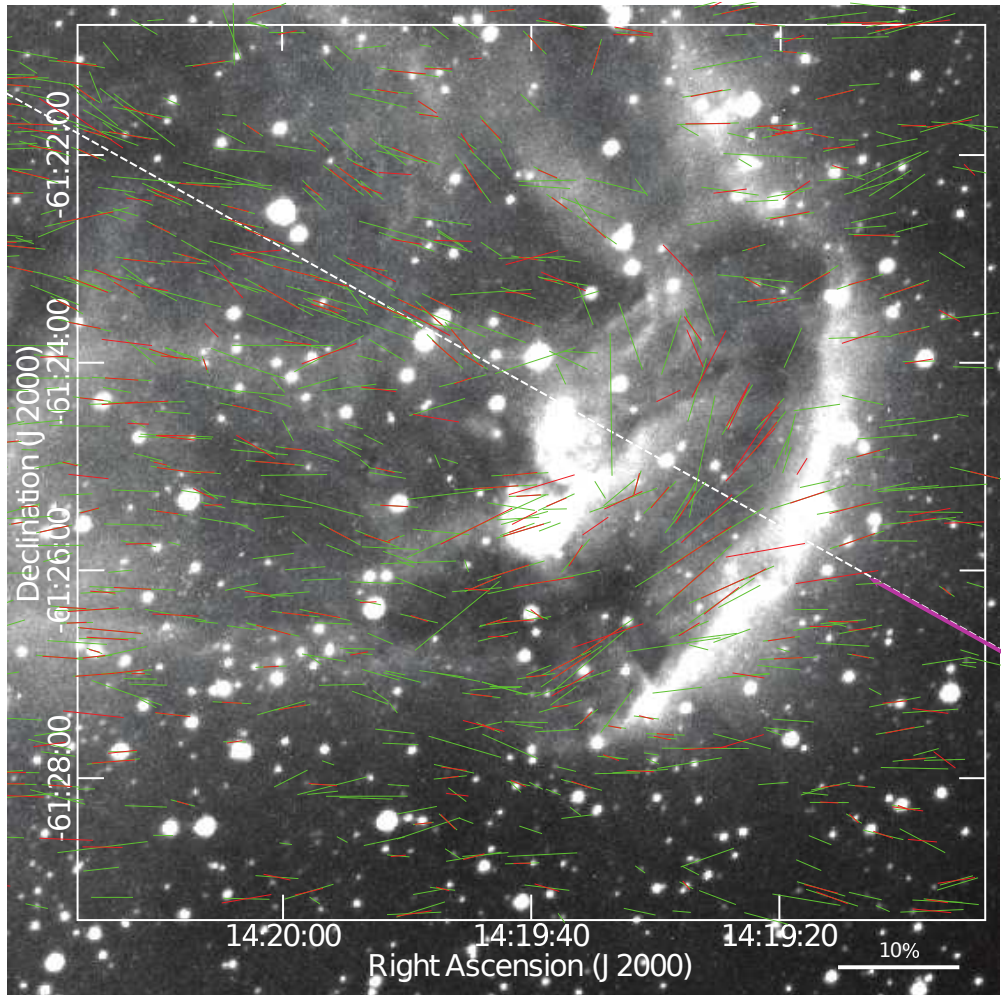}
\caption{
Polarization vector map superposed on the SuperCOSMOS H$_\alpha$ image.
$H$- and $K_{\rm s}$-band polarization vectors are shown with green and red vectors, respectively.
A 10\% vector is shown at the right bottom. 
The incident UV radiation direction is shown with a purple arrow, 
and its extended line (our adopted approximately symmetry axis) is shown by a white dashed line. 
Nearly all of the vector directions are consistent between two bands within their errors of $\la10\degr$ for sources with polarization vectors both at two bands. 
}
\label{HK-pol}
\end{figure}

\begin{figure}
\epsscale{0.5}
\plotone{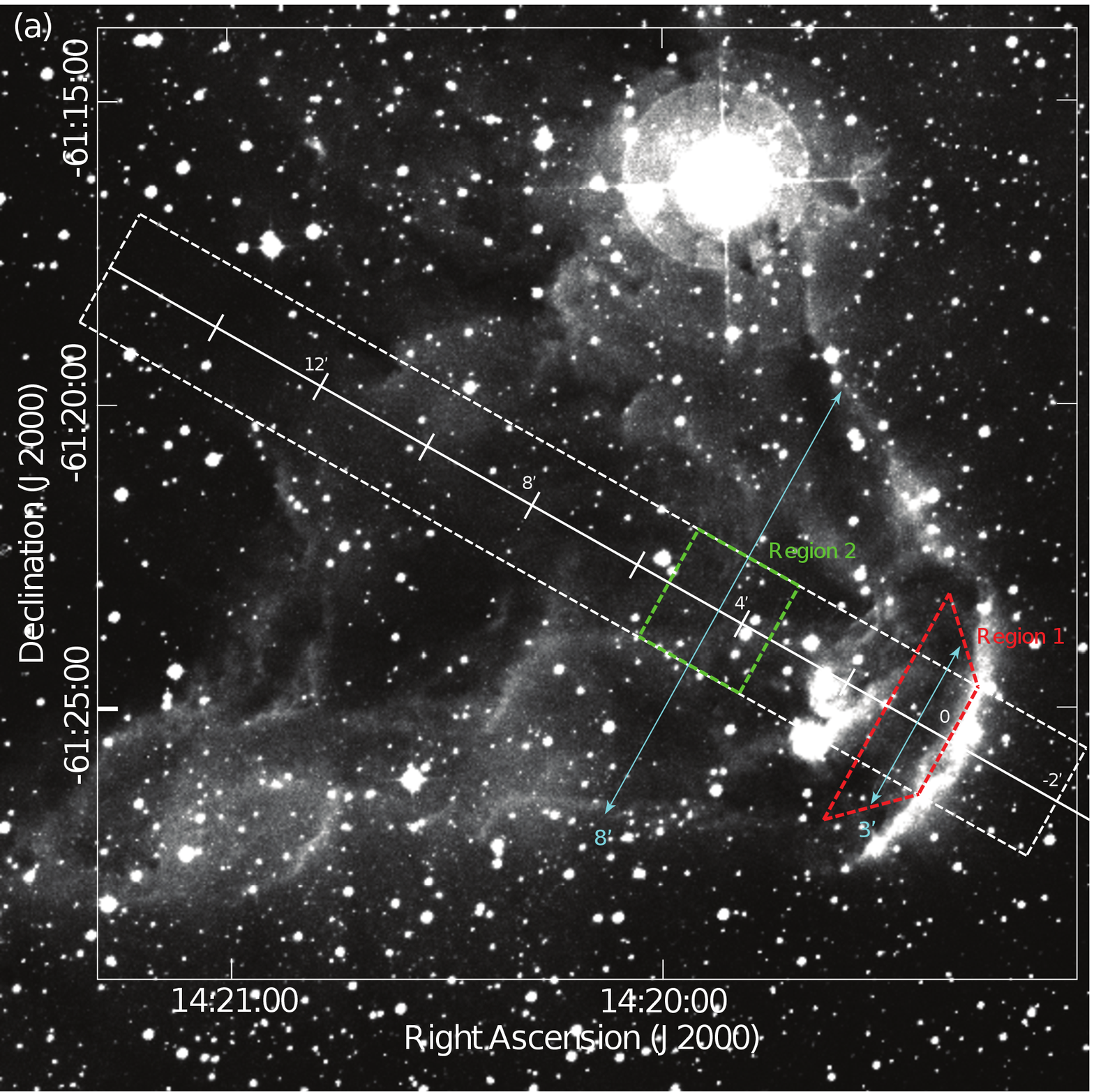}
\plotone{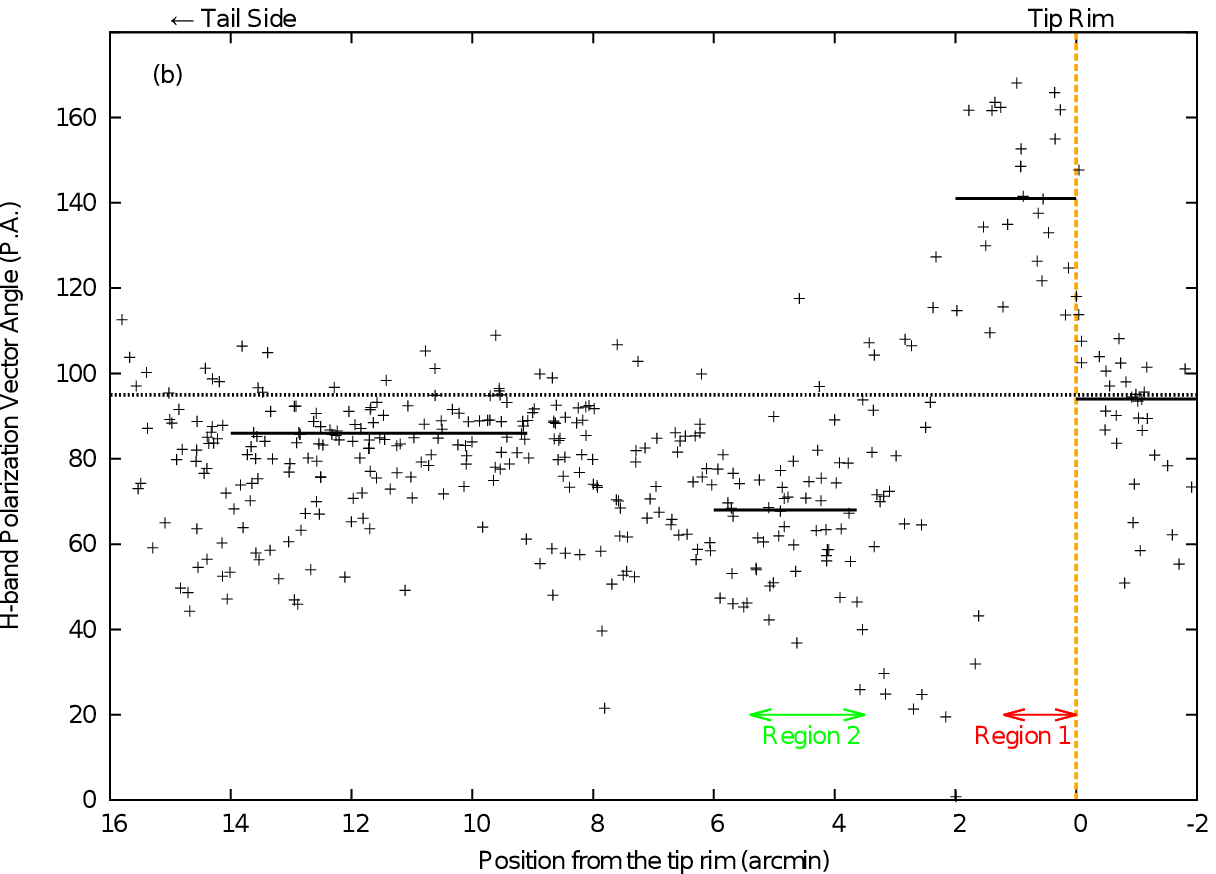}
\caption{
(a) 
Areas for analysis superposed on the SuperCOSMOS H$_\alpha$ image. 
A white dashed box with a width of 2$\arcmin$ is the area that contains the sources used for plotting (b). 
Region~1 (red, dashed trapezoid) and Region~2 (green, dashed rectangle) are the areas that we used for analysis. 
Widths, 
which we adopted as the depths of two regions for analysis, 
are shown with cyan arrows.
(b) 
Polarization angle of $H$-band vectors versus position from the tip bright rim. 
A dashed line shows the average of the ambient polarization direction. 
Solid lines show the averages of the polarization directions and the sample ranges (see Section 3.2). 
}
\label{fig7}
\end{figure}

\begin{figure}
\epsscale{0.32}
\plotone{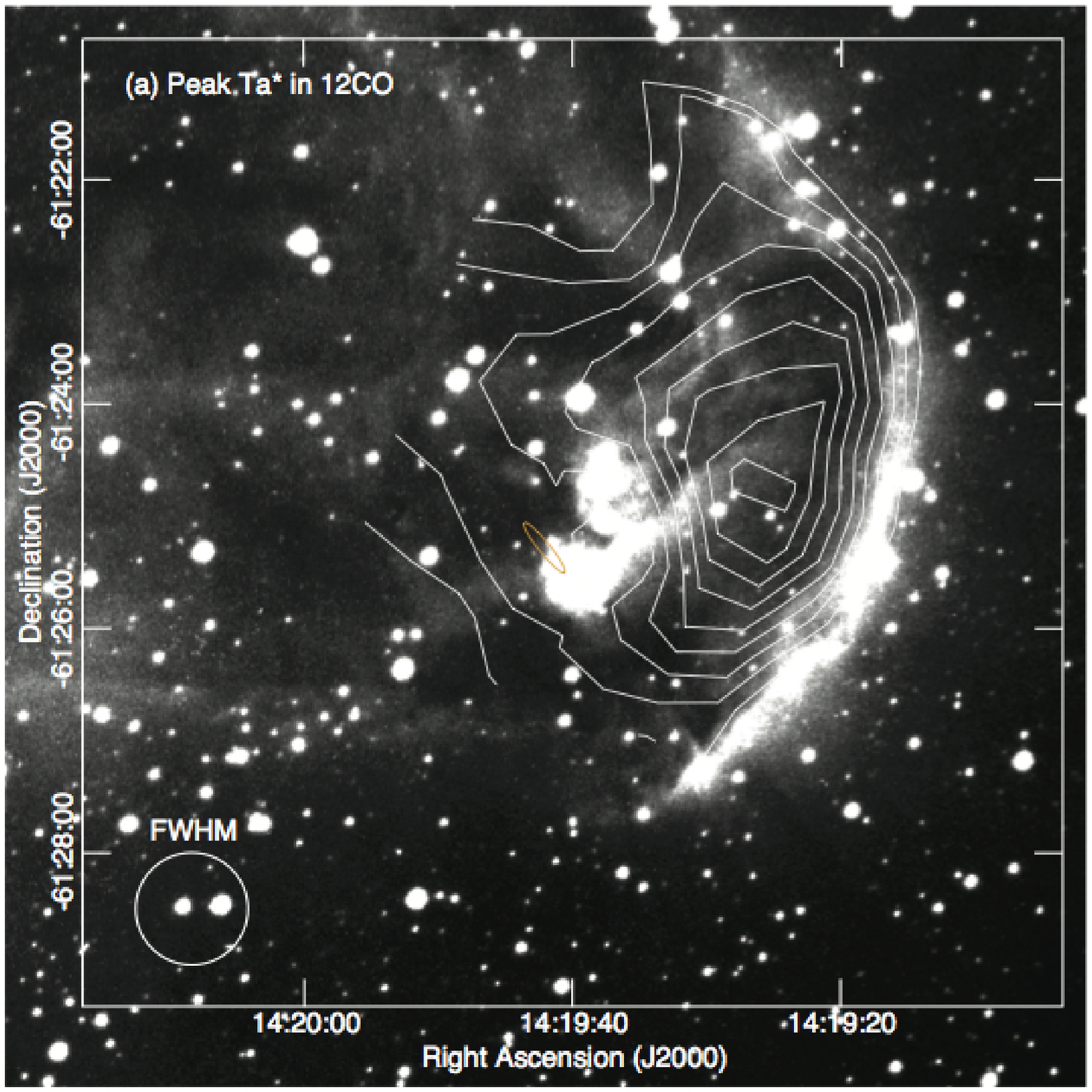}
\plotone{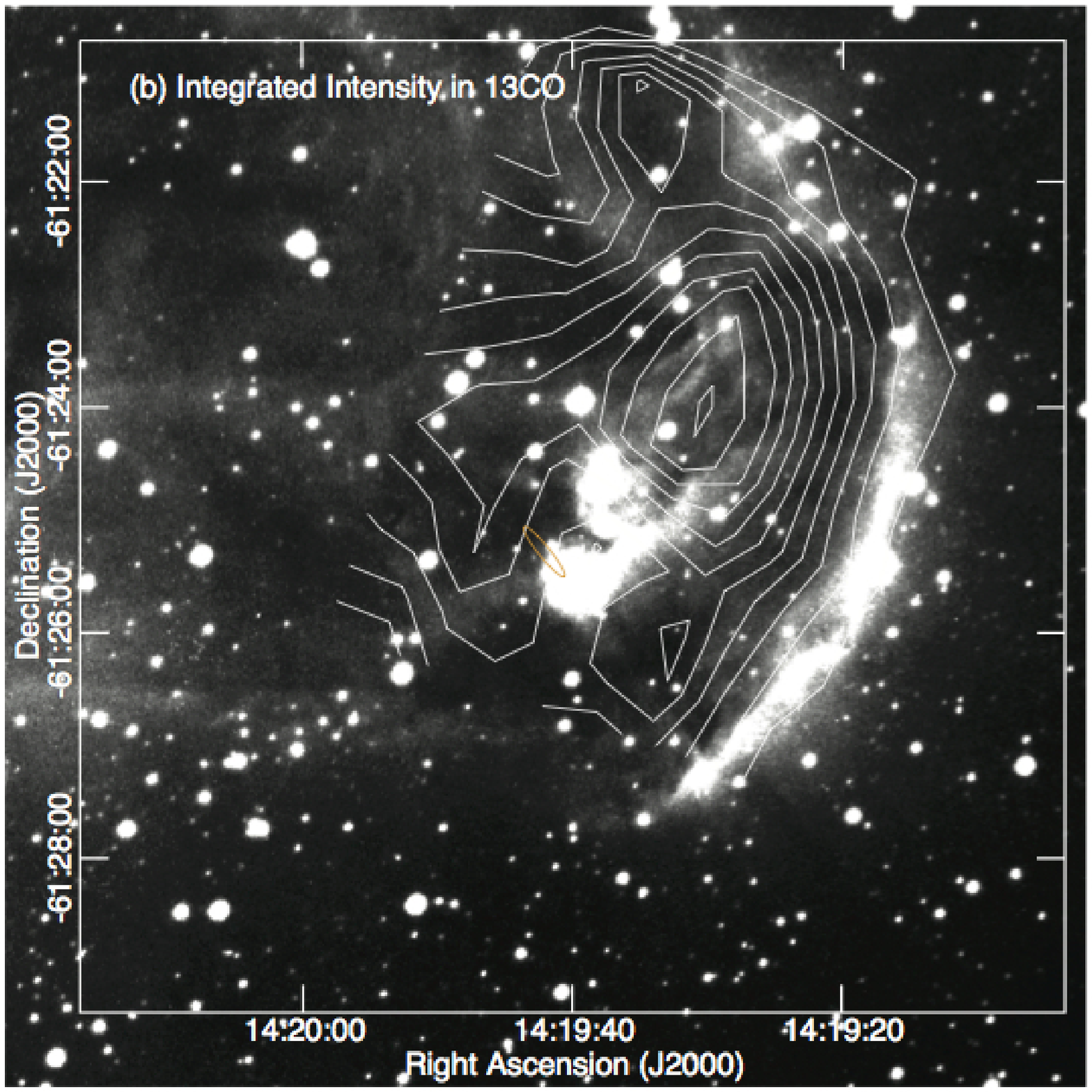}
\plotone{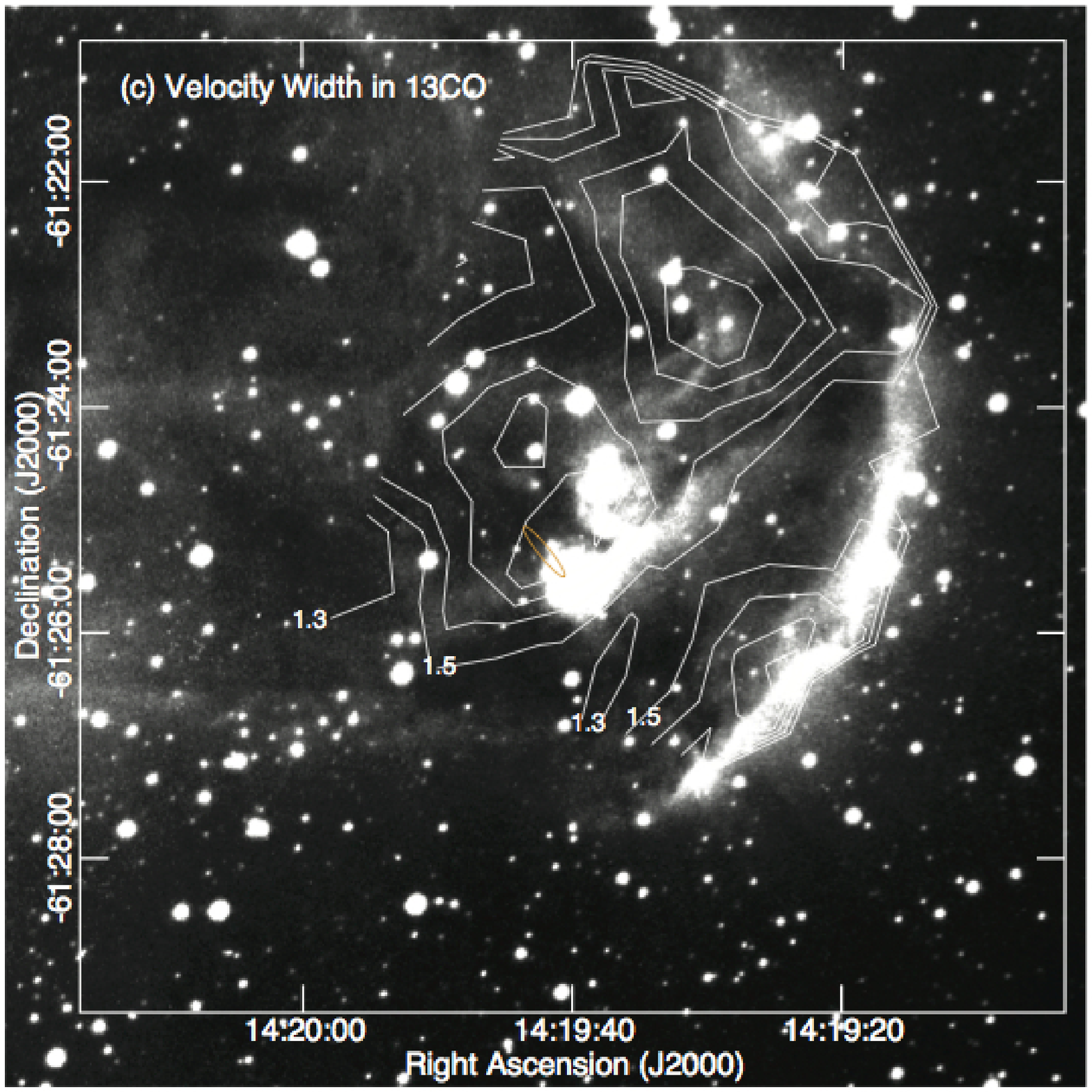}
\caption{
(a) 
Peak $T_{\rm A}^*$ map of $^{12}$CO (1--0) superposed on the SuperCOSMOS H$\alpha$ image. 
The lowest contour level is 10 K, and the contour interval is 2 K in $T_{\rm A}^*$. 
The beam size is shown in the lower left corner. 
(b) 
Integrated intensity map of $^{13}$CO (1--0) superposed on the SuperCOSMOS H$\alpha$ image. 
The velocity range is $-31.4$ to $-25.9$ km s$^{-1}$ in $v_{\rm LSR}$. 
The lowest contour level is the $6\sigma$ level ($=1.04$ K km s$^{-1}$ in $T_{\rm A}^*$), 
and the contour interval is $6\sigma$. 
(c) Velocity width map of $^{13}$CO. 
The lowest contour level is 1.3 km s$^{-1}$, and 
the contour interval is 0.2 km s$^{-1}$. 
The position of IRAS 14159-6111 is shown with an orange error ellipse of position. 
}
\label{fig9}
\end{figure}

\begin{figure}
\epsscale{1}
\plotone{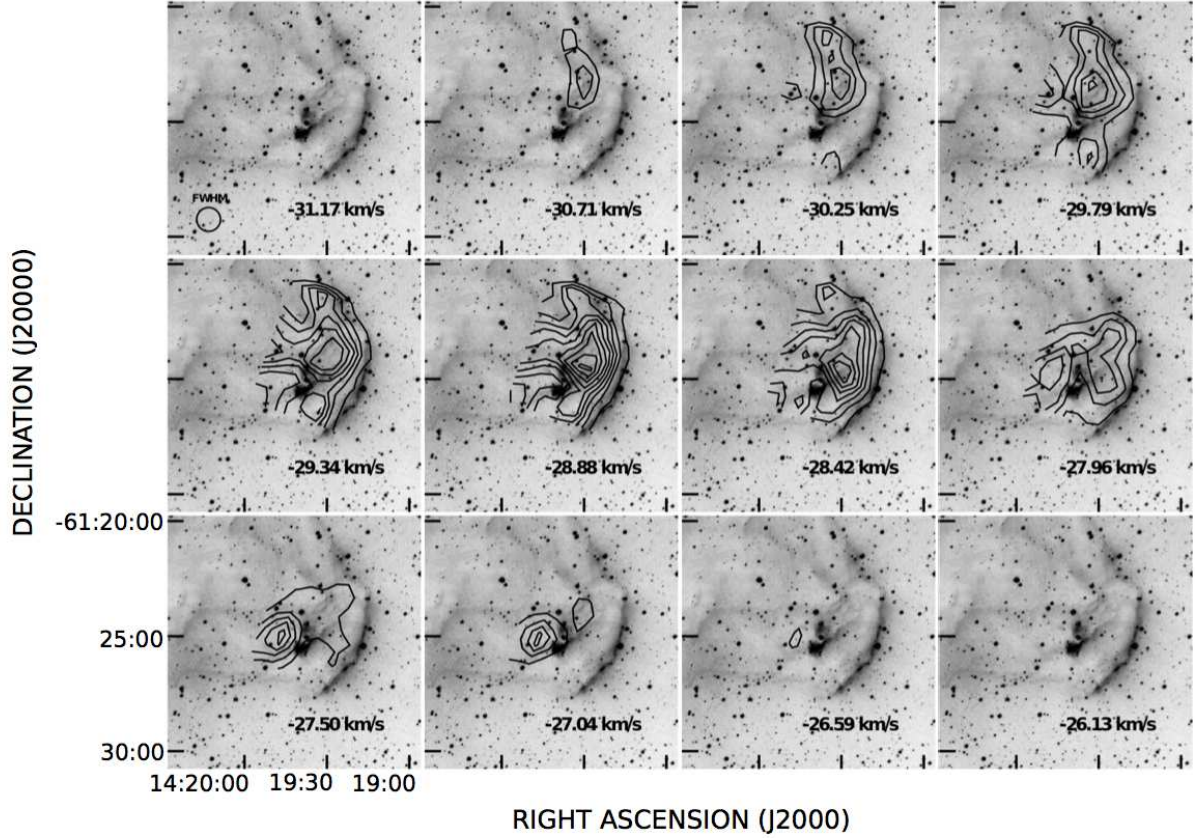}
\caption{
Channel maps of $^{13}$CO superposed on the SuperCOSMOS H$_\alpha$ image. 
The center LSR velocity of each channel is shown on the bottom right of each panel. 
The lowest contour level is the $6\sigma$ level ($=0.3$ K km s$^{-1}$ in $T_{\rm A}^*$), and the contour interval is 6$\sigma$. 
The beam size is shown in the lower left corner in the panel of $v_{LSR}=-31.17$ km s$^{-1}$. 
}
\label{fig10}
\end{figure}

We try to estimate the physical quantities related to the magnetic fields in the two regions; 
one is the tip region (Region~1) just behind the tip bright rim, 
and the other is the inside region (Region~2) on the cloud symmetric axis. 
Region~1 and Region~2 are shown by a red trapezoid and a green square, respectively on Figure \ref{fig7}a. 
Estimating the magnetic field strength with the CF \citep{chandra53} method requires an area 
where the magnetic field is uniform.  
Since the polarization vectors of Region~1 appear to uniformly follow the curve of the tip bright rim (Figure \ref{HK-pol}), 
we selected Region~1 that is considered to be the compressed layer by RDI as discussed in Section 4.1. 
Since the magnetic field of Region~2 seems to be rather uniform as shown in Section 3.2,  
we selected Region~2 so that its area is the same as that of Region~1, 
well outside Cloud~B.  
The southwest half of Region~2 is covered by the CO observations, 
and we can crudely  evaluate the velocity dispersion of Region~2. 
Here, 
we use only the sources with $H$ and/or $K_{\rm s}$ vectors on Figure \ref{HK-pol}. 

For Region~1, 
we adopt a thick, uniform disk approximation and assume that we view this disk edge-on. 
In this case, 
the density can be obtained by dividing the central/maximum column density 
toward the axis by the diameter/width of the disk, 
i.e., the depth. 
The average of color excess $E(H-K_{\rm s})$ is $\sim1.5\pm0.7$ mag for 20 sources detected 
toward the central part of Region~1 (within 60$\arcsec$ of the symmetric axis), and 
$N_{{\rm H}_2} \sim2.9\times10^{22}$ cm$^{-2}$ is roughly obtained, 
using the reddening law, 
$E(H-K_{\rm s})=0.065 \times A_{\rm V}$ \citep{chini98} and the standard gas-to-extinction ratio, 
$N_{{\rm H}_2}/A_{\rm V} \sim 1.25 \times 10^{21}$ cm$^{-2}$ mag$^{-1}$ \citep{dick78b}. 
This estimated column density seems to be consistent with that estimated from the CO data, 
$\sim$ 1.3-4.1 $\times10^{22}$ cm$^{-2}$. 
Adopting a width of $\sim$3$\farcm0$ ($\sim$1.4 pc) as the depth of Region~1, 
we can derive a number density of $n_{{\rm H}_2}\sim6.5\times10^3$ cm$^{-3}$. 
Since Region~2 is located around the symmetric axis of the cloud, 
the column density in Region~2 is expected to be close to the central/maximum column density 
if the local density is uniform. 
The average of color excess $E(H-K_{\rm s})$ is $\sim1.1\pm0.4$ mag for 41 sources toward Region~2 and 
$N_{{\rm H}_2} \sim2.1\times10^{22}$ cm$^{-2}$ is roughly obtained. 
This column density also seems to be consistent with that estimated from the CO data, $\lesssim2.1\times10^{22}$ cm$^{-2}$. 
Adopting the width of $\sim$8$\farcm0$ ($\sim$3.5 pc) as the depth of Region~2, 
$n_{{\rm H}_2}\sim2.0\times10^3$ cm$^{-3}$ can be obtained in Region~2. 
The estimated number density in Region~1 is $\sim$ 3-4 times higher than that in Region~2, 
supporting the idea that the Region 1 was strongly compressed by UV and then became higher density. 
This result does not seem to match the simulation results of the non-magnetized clouds 
that the number density in the tip is expected to be by more than one order of magnitude higher 
than that of the area away from the tip \citep[e.g. Figures 3, 7 and 10 of][]{miao09}. 
However, 
the simulation taking into account the magnetic pressure predicts a smaller difference 
of a factor of $<$ 10 \citep[e.g. Firgure 8 of][]{miao06}, 
which is more consistent with our observations. 
\citet{moto13} also suggests flatter column density profiles in the magnetized cases than in the non-magnetized ones.

We estimated the strengths of the magnetic fields ($B_\|$) projected on the plane of the sky toward the two regions, 
using the CF method, 
$$B_\|=Q\sqrt{4\pi\rho}\frac{\sigma_v}{\sigma_\theta}$$
where $\rho$ is the mean density of the cloud, 
$\sigma_v$ is the velocity dispersion,  
$\sigma_\theta$ is the dispersion of the polarization vector angles, and
 $Q\sim$0.5 is a correction factor for $\sigma_\theta\lesssim25\degr$ \citep{ostri01}.  
The magnetic field toward Region~1 seems to be slightly bent along the tip bright rim, 
so we modeled the shape of the magnetic field with a parabolic function, 
assuming that the magnetic field direction is uniform along the tip rim.  
The magnetic field was adequately fitted with the slopes of the polarization vectors 
by rotating the parabolic function by 60$\degr$ in an easterly direction. 
We excluded sources having a large difference from the average. 
If the vectors are shown at two bands for a single source in Figure \ref{HK-pol}, 
only the single position angle with less error was used for the fitting. 
At first, 
we calculated the dispersion ($\sigma_{\theta}$) from the best-fit model for 23 sources in Region~1. 
Then, 
removing the dispersion due to the measurement uncertainties of the polarization angles, 
we obtained the corrected dispersion $\sigma^\prime_{\theta}$ of $\sim$11$\fdg4$ ($\sigma^\prime_{\theta}\sim\sqrt{\sigma_{\theta}^2-\frac{\sum \Delta\theta^2}{n}}$),  
and used this for estimating the magnetic field strength with the CF method. 
The $^{13}$CO velocity widths toward Region~1 are $\sim$ 1.3-1.5 km s$^{-1}$ in Figure \ref{fig9}c.
Therefore, 
we adopted 1.4 km s$^{-1}$ as a representative velocity width toward Region 1, 
which corresponds to the velocity dispersion ($\sigma_{\upsilon}$) $\sim$0.64 km s$^{-1}$. 
Adopting the molecular weight $\mu\sim$2.3, 
we obtained $B_\|\sim$90 $\mu{\rm G}$. 

For Region~2, 
assuming that the direction of the magnetic field is uniform nearly along the symmetric axis, 
we obtained $\sigma^\prime_{\theta}$ $\sim$16$\fdg4$ for 41 sources in Region~2 after removing the uncertainties. 
The CO observations only covered the southwest half of Region~2. 
The  $^{13}$CO velocity width toward the southwest half of Region~2 appears to be $\sim$ 1.4-1.9 km s$^{-1}$, 
and to decrease with distance away from the southwest edge of Region~2 along the symmetry axis.  
Therefore, 
we adopted the velocity width of $\sim$1.4 km s$^{-1}$ near the center of Region~2 as a representative velocity width of Region~2, 
and obtained $B_\|\sim$30 $\mu{\rm G}$. 

It should be noted 
that the polarization vector maps toward SFO~74 might be contaminated by the polarized sources lying foreground and/or background to the cloud. 
However, 
this effect is probably small toward the dense parts of the cloud, 
i.e., 
Region~1 and 2, 
because 
1) the sources far behind the cloud, 
which have already undergone large interstellar absorption at the far side of the cloud, 
would not be detectable with our criteria, 
and 
2) the extinction of foreground dwarfs, 
which are located at $H-K_{\rm s}\la0.2$ (Figure 3), 
is estimated to be $A_V$ $\la$1.5 mag, 
adopting $\sim$0.1 mag as the intrinsic color ($H-K_{\rm s}$) of the dwarfs and using the reddening law, $E(H-K_{\rm s})=0.065 \times A_V$ \citep{chini98}.
This extinction is much smaller than those of Region 1 and 2 ($A_V\sim$23 and 17 mag). 
Therefore, 
the contamination from the foreground and/or background interstellar polarization is negligibly small in Region~1 and 2. 

The magnetic field of Region~2 might be also enhanced by the convergence flow with respect to the original magnetic field, 
while that of Region~1 might be enhanced by the ionized front. 
The ratio of the magnetic field strength of Region~1 to that of Region~2 is derived to be $\gtrsim$2, 
implying that 
the magnetic field of Region~1 has been more strongly enhanced. 
The CF method is considered to be a rough estimate of the magnetic field strength, and 
the derived strength might have a large uncertainty. 
However, 
we can still discuss the relative ratio of the derived magnetic field strengths because 
the physical parameters used were derived in the same way.  
The uncertainty in density is the largest, 
but we note that the magnetic field strength is proportional to the root of the density. 

We compare the magnetic pressure, 
$P_B=B^2/8\pi$, 
with the turbulent pressure, 
$P_{\rm turb}=\rho\sigma_{\rm turb}^2$ in Region~1.
$P_B$ is calculated to be $\sim3.3\times10^{-10}$ dyn cm$^{-2}$, 
adopting $B_\|\sim$90 $\mu{\rm G}$ as $B$.
This estimated value would be the lower limit 
because the magnetic field may not be running perpendicular to the line of the sight. 
Using 1-D velocity dispersion ($\sigma_{\upsilon}$) of $\sim$0.64 km s$^{-1}$, 
$P_{\rm turb}$ is 
$\sim3.1\times10^{-10}$ dyn cm$^{-2}$. 
These comparable estimated values imply that 
the magnetic pressure significantly contributes to the internal pressure,  and that 
the total pressure is $\sim1.0\times10^{-9}$ dyn cm$^{-2}$.  
For Region~2, 
we obtained $P_B\sim4.7\times10^{-11}$ dyn cm$^{-2}$ and $P_{\rm turb}\sim9.2\times10^{-11}$ dyn cm$^{-2}$.
The magnetic pressure is estimated to be slightly smaller than the turbulent pressure in Region~2. 

Then, 
we compare the total internal pressure in Region~1 with the external pressure, 
i.e., 
the radiation pressure $P_{\rm rad}$ and the post shock pressure by the ionization front. 
The maximum ionizing flux is 
$F=4.5\times10^{8}$ cm$^{-2}$ s$^{-1}$ 
and the energy of an ionizing photon $h\nu\ge$13.6 eV. 
If we assume the maximum average energy of  an ionizing photon 
$h\nu_{\rm max}=20$ eV (in fact it is much lower than that), 
then we can find the upper limit of $P_{\rm rad}=F/c<4.8\times10^{-13}$ dyn cm$^{-2}$. 
It is seen that $P_{\rm rad}$ is negligible. 
To estimate the post shock pressure, 
we assumed plausible physical quantities of $n\sim$ a few $\times10^{3}$ cm$^{-3}$ and $T\sim$8000 K
for the number density and the temperature of the ionization front, respectively. 
We thus derived the external pressure of a few $\times$ 10$^{-9}$ dyn cm$^{-2}$. 
This value is comparable to or slightly larger than the total internal pressure, 
and thus 
these two regions may be in an equilibrium state or 
in a weak overpressure state of the ionization front. 
If this is the case, 
we suggest that the magnetic field plays an important role in the pressure balance between these two regions and in the evolution of Cloud~A. 

The dynamical state of a magnetized cloud is measured by the ratio between the cloud mass ($M_{\rm cloud}$) and the magnetic flux ($\Psi$), 
i.e., the mass-to-flux ratio, which is given by $M_{\rm cloud}/\Psi=\mu m_{\rm H}N_{\rm H_2}/B\sim2.0\times(M_{\rm cloud}/\Psi)_{\rm critical}$ 
for Region 1, where $(M_{\rm cloud}/\Psi)_{\rm critical}$ is the critical value for magnetic stability of the cloud 
\citep[$=(4\pi^2G)^{-1/2}$;][]{nakano78}. 
Here, we assume that the magnetic field is almost perpendicular to the line of sight,
but that is not necessarily so.
Therefore, 
the actual magnetic field strength might be larger, and 
the net mass-to-flux ratios could be close to the critical value.
This implies that the magnetic field is likely to play an important role in the cloud dynamics, 
such as the evolutional time scale of the cloud, and the density and the thickness of the compressed layer, 
and thus star formation in Cloud~A.
For Region~2, 
$M_{\rm cloud}/\Psi$ is given by $\sim3.8\times(M_{\rm cloud}/\Psi)_{\rm critical}$, 
which is by a factor of $\sim$2 larger than that in Region~1, 
and the turbulent pressure is dominant. 

\subsection{Flat-Topped Shape}
As seen in Figure \ref{J-pol}, 
the bright rim of SFO~74 appears to consist of two parts; 
the tip facing the exciting star and the wings extending toward both tail sides (Figure \ref{fig12}).
We briefly consider this flat-topped shape.

The magnetic fields just behind the wing rims seem to run along the bright rims like that just behind the tip rim (Figure \ref{J-pol}).
The strength of magnetic field just behind the tip, however, 
could be larger than that just behind the wing. 
The layer just behind the tip rim is perpendicular to the incident UV direction and 
might be more powerfully compressed by the UV radiation than that just behind the wing rims,
which are nearly parallel or oblique to the UV radiation. 
We speculate that this flat top shape is formed by the larger increase of the magnetic field strength just behind the tip. 

As for the physical explanation of the flat-topped shape of SFO~74, 
we look at the magnetohydrodynamic equation of motion, 
which is one of the four astrophysical fluid equations,
\[
 \rho \frac{d\bf{V}}{dt} = - {\bf \bigtriangledown} p + {\bf \bigtriangledown}  \cdot  \bf{T},
 \label{mag}
\]
where all of symbols are of their conventional physical meaning, 
and the magnetic pressure tensor $ T_{\mathrm{ij}} = \frac{B_i B_j - \delta_{i j} B^2 / 2}{4 \pi}$ ($i,j=1,2,3$). 

According Figure \ref{fig7}b, 
the average magnetic field direction behind the tip has a P.A. of $\sim$68$\degr$, 
approximately parallel to the central symmetric line of SFO~74 and the UV radiation direction ($\sim$60$\degr$). 
We choose $z$-axis is parallel the direction of the average magnetic field and UV radiation, 
and assume the field just behind the compressed surface layer of SFO~74, 
the above equation becomes:
\[
 \rho \frac{d\bf{V}}{dt} = - {\bf \bigtriangledown \cdot P }
 \label{pressure}
\] 
where the deduced pressure tensor,  
\[ 
{\bf P} = \left (
\begin{array}{ccc}
 p + \frac{B^2}{8 \pi } & 0 & 0 \\
 0   &  p + \frac{B^2}{8 \pi } & 0   \\
 0  &  0 & p - \frac{B^2}{8 \pi }
\end{array} 
\right ). 
\]

We can see that the magnetic field increases the cloud pressure by an amount of $\frac{B^2}{8 \pi}$ 
in the elongation direction perpendicular to the magnetic field ($x$ and $y$), 
but decreases the 
cloud pressure by the same amount in the parallel direction ($z$) \citep{cai85}. 
Therefore, we can expect that the $z$ axis directional magnetic field B in SFO~74 makes 
the structure of the cloud stretch in the width direction perpendicular to the magnetic field line and contract in the parallel length direction. 
This effect further quashes the magnetic field line loops created by UV radiation induced shock, 
so that the loop lines looks around the bright rim. 
The strength of this magnetic effect increases with the local density due to a positive dependence of magnetic field B on number density. 
Therefore a flatter bright tip rim than that in zero magnetic field case appears on the star facing side of SFO~74. 

Although our main focus in this paper is to investigate the physical properties of the head part of SFO~74, 
we are also interested in deriving a reasonable picture on the development of the morphology of the two wing rim structure shown in Figure \ref{fig12}. 
As we analyzed above, the magnetic field line behind the Region 1 part of the type B SFO~74 is much less effected by the shock, 
increase of the magnetic pressure along x-y direction expands the width of the tail section (Region~2) more and more efficiently toward the rear end of SFO~74 than in the head part, 
so that we see the gradually outward curved wing rim,
and the magnetic field lines naturally parallel to the wing rim. 
However, 
the wildly curved-out part at the rear end of the wing rim might not be caused by the expansion effect of the z-directional magnetic field only, 
which might also be the result of the Rocket effect induced by the photo-evaporation of the ionized gas \citep{oort55}. 
In order to reveal the actually morphological and physical evolution of SFO~74, 
we are preparing a 3-D simulation by fully treating the magnetic field (Miao et al. 2014, in preparation).

\subsection{Star Formation Activity}

\subsubsection{Cloud~A} 
\begin{figure}
\epsscale{0.45}
\plotone{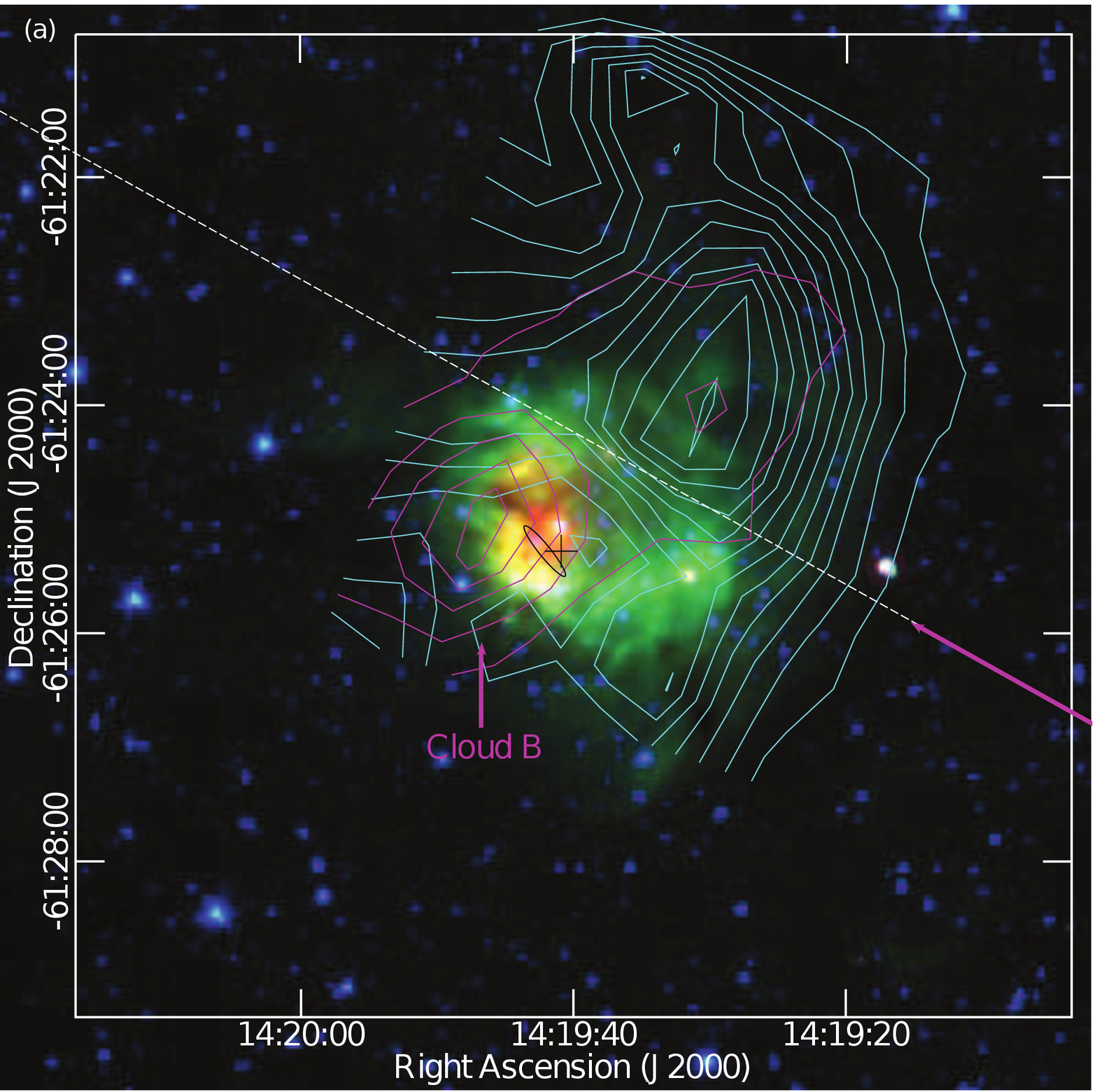}
\plotone{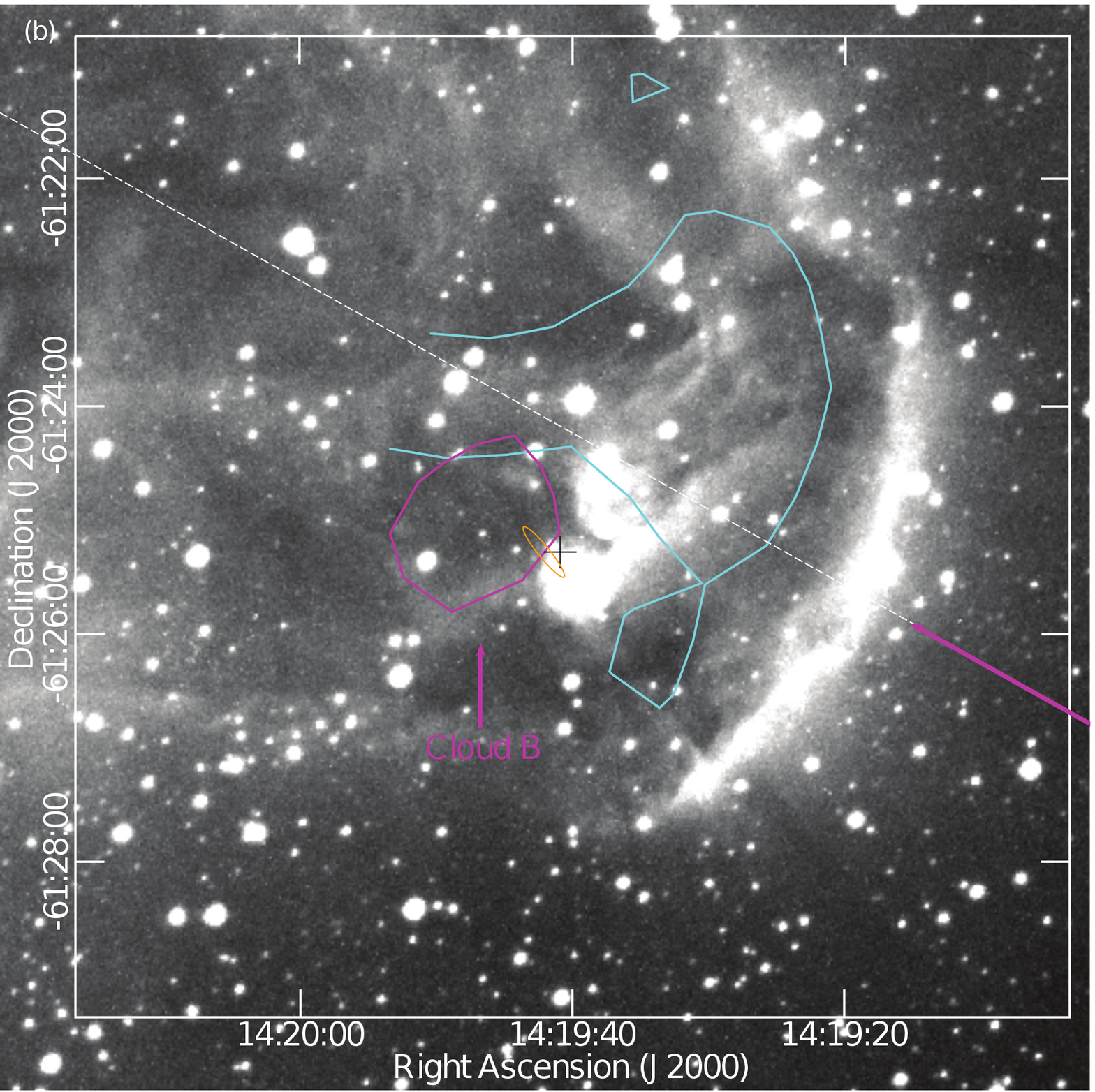}
\caption{
(a) Integrated intensity maps of Cloud~A (cyan; $v_{\rm LSR}=-31.35$ to $-27.78$ km s$^{-1}$) 
and of Cloud B (magenta; $v_{\rm LSR}=-27.69$ to $-25.94$ km s$^{-1}$) superposed on a three color composite image, 
which has been constructed from the Spitzer-GLIMPSE 3.5 $\mu$m (blue) image, 8 $\mu$m (green) image \citep{ben03} and Spitzer-MIPSGAL 24 $\mu$m (red) image \citep{carey09}. 
The lowest contour level for Cloud~A is the $6\sigma$ level ($=0.85$ K  km s$^{-1}$ in $T_{\rm A}^*$), and the contour interval is $6\sigma$. 
The lowest contour level for Cloud~B is the $6\sigma$ level ($=0.6$ K km s$^{-1}$ in $T_{\rm A}^*$), and the contour interval is $6\sigma$. 
(b) Cloud~A and B positions superposed on the Super COSMOS H$_\alpha$ image. 
The solid lines show 50\% of the peak integrated intensity in $^{13}$CO.   
The cyan line is for Cloud A, and the magenta one for Cloud~B. 
The position of IRAS 14159-6111 is shown with a black or orange error ellipse of position. 
The peak position of radio continuum with ACTA \citep{thom04} is shown by a black cross. 
The symmetric axis is shown by a white dashed line. 
}
\label{fig11}
\end{figure}

\begin{figure}
\epsscale{0.5}
\plotone{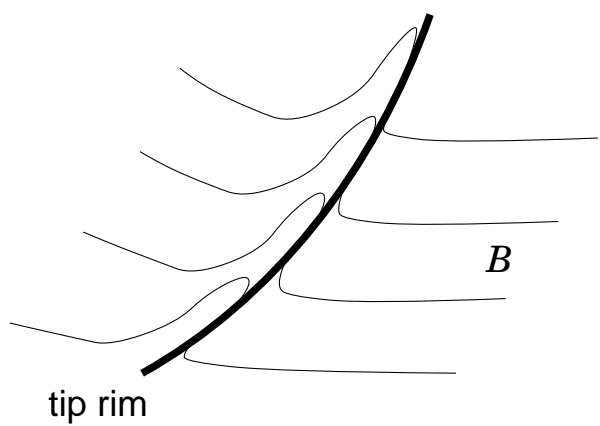}
\caption{
Schematic drawing of the magnetic field just behind tip bright rim. 
}
\label{MFC}
\end{figure}

\begin{figure}
\epsscale{0.5}
\plotone{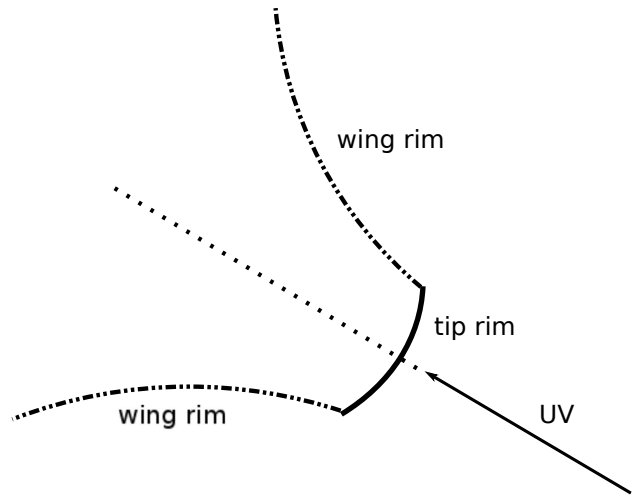}
\caption{Schematic drawing that shows the flat-topped shape of SFO 74. 
The bright rim seems to consist of two parts; tip rim and wing rims. 
}
\label{fig12}
\end{figure}

As shown in Figure \ref{fig11}, 
the peak position of radio continuum with ACTA \citep{thom04} and the position of IRAS/MSX point source are not located toward Cloud~A, but Cloud B. 
We did not see any sign of high-mass star formation from the Spitzer composite image (Figure \ref{fig11}a), 
and of $^{12}$CO molecular outflow from the Mopra data. 
Thus, 
the star formation in Cloud A may not be currently active. 
The 8 $\micron$ emission is considered to be dominated by PAHs and useful for identifying a cavity wall of {an} HII region \citep[e.g.,][]{rath02,urqu03,urqu09}. 
No strong 8 $\micron$ emission is associated with Cloud~A (see green color in Figure \ref{fig11}a), 
and the $^{12}$CO data suggests the external heating from the exciting star of RCW~85.  
Thus, 
we concluded that 
Cloud~A was not strongly affected by the UCHII region. 

\subsubsection{Cloud B}
In contrast to Cloud~A, 
Cloud~B has some signs of active star formation. 
So, 
we examine more closely star formation in Cloud~B. 

\begin{figure}
\epsscale{0.4}
\plotone{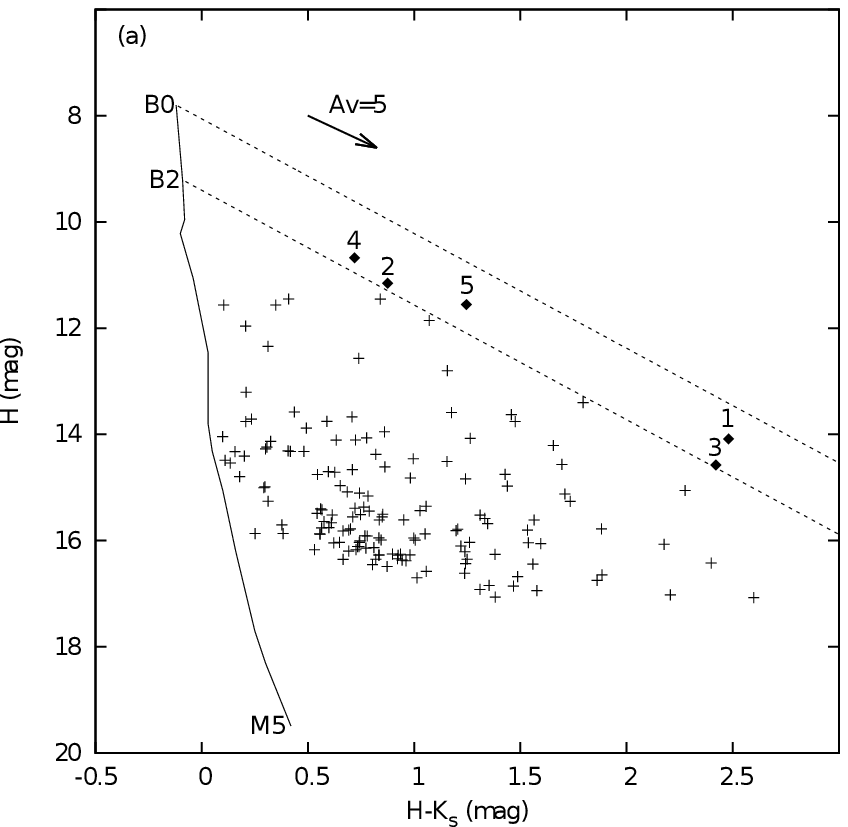}
\plotone{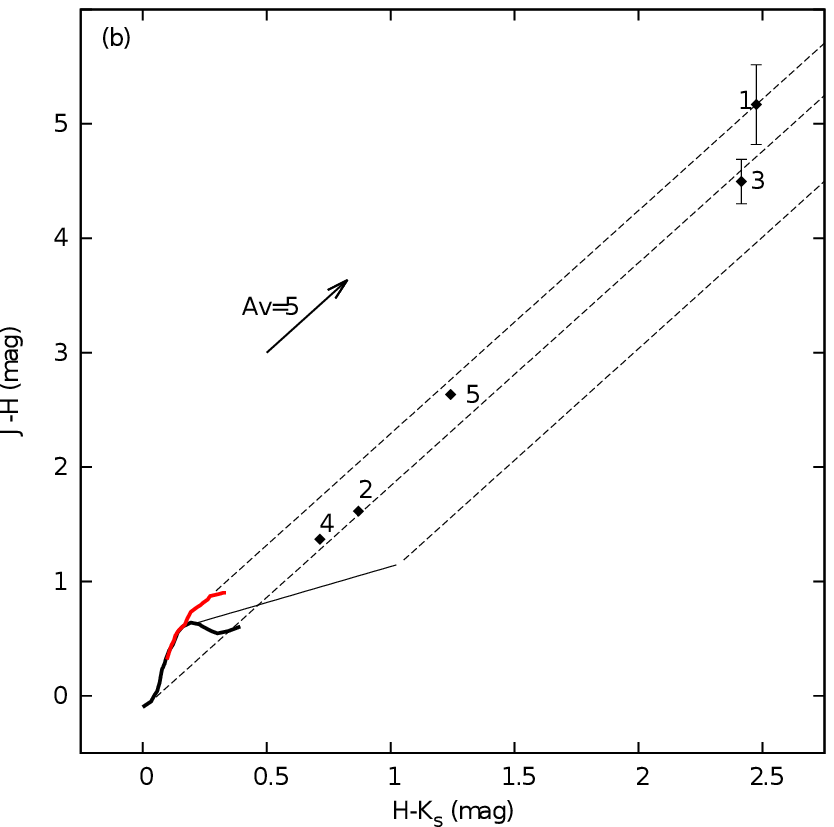}
\epsscale{0.9}
\plottwo{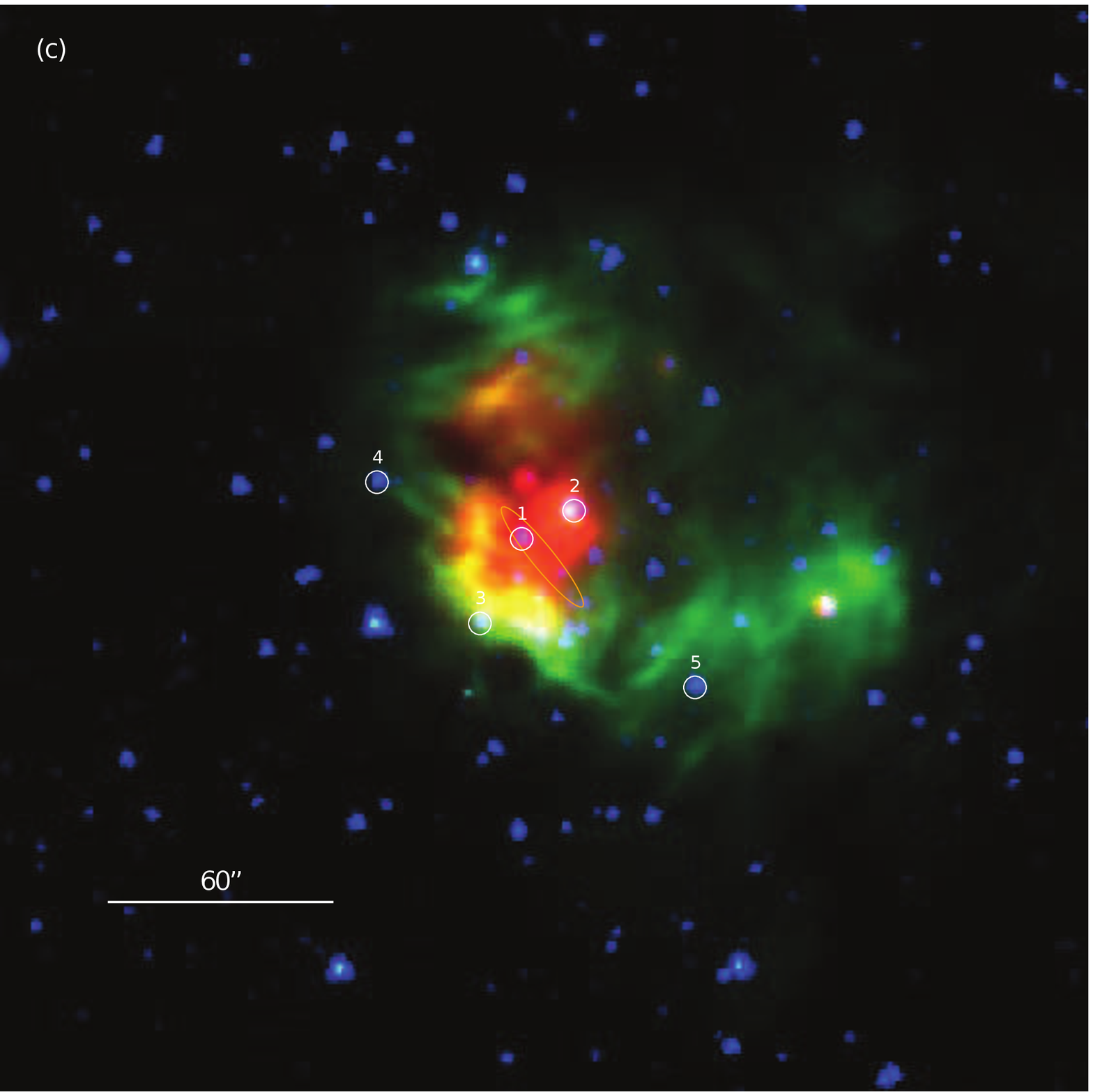}{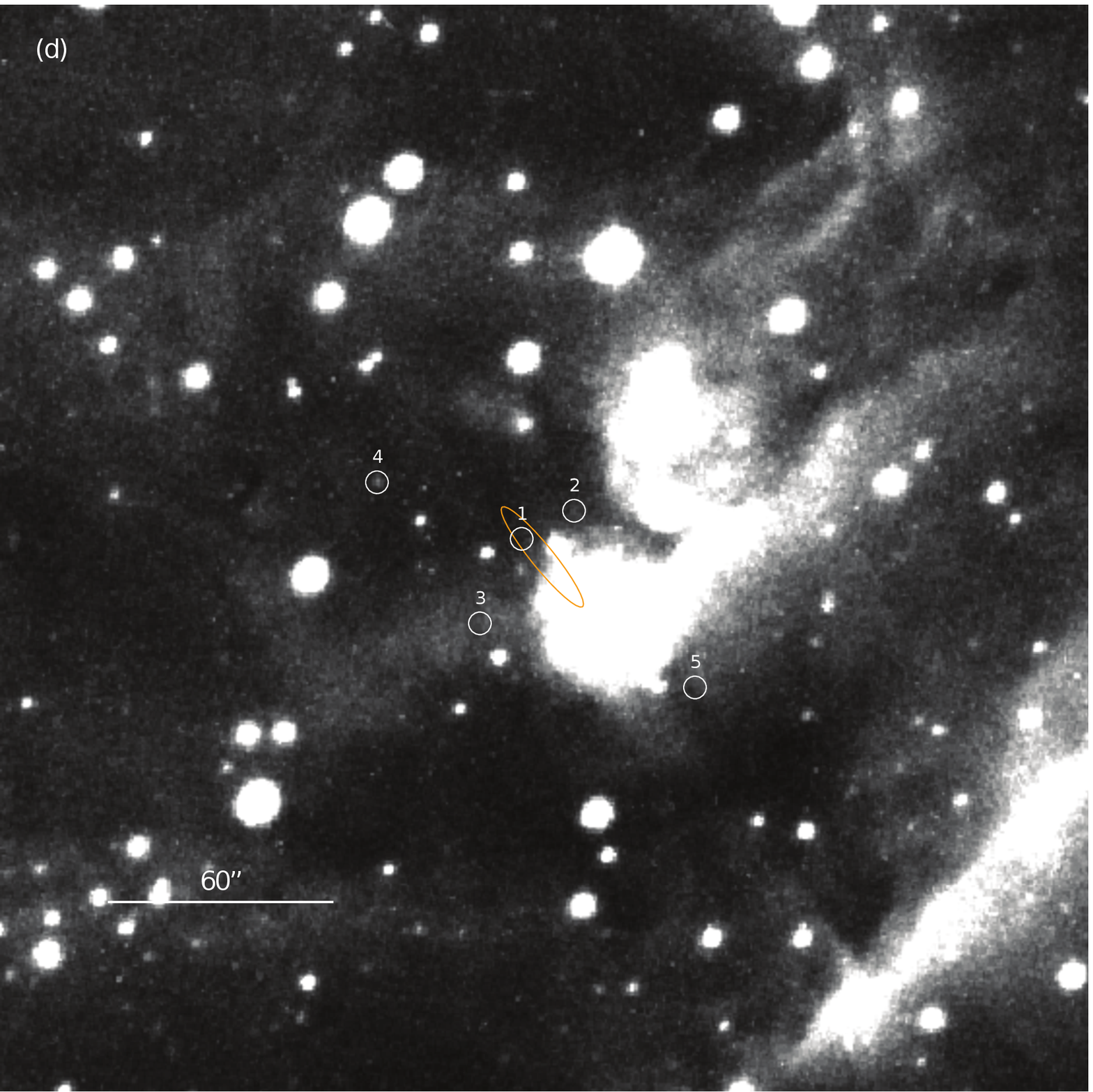}
\caption{\footnotesize (a) 
$H$ versus $H-K_{\rm s}$ diagram for the near-IR sources within 60$\arcsec$ of the IRAS position. 
Numbered sources are candidates for the exciting star of UCHII region.  
The locus of main-sequence stars at a distance of 1.5 kpc is shown by solid line \citep{stahler04}.  
To obtain the $H$ magnitude and $H-K_{\rm s}$ color of the main sequence star, 
Table~A5 of \citet{keny95} and the transformation equations of the 2MASS explanatory supplement  were used. 
(b) 
$J-H$ vs. $H-K_{\rm s}$ color-color diagram for the candidates of the exciting star. 
This diagram is created in the exactly same way as Figure \ref{fig2}. 
Because sources \#1 and \#3 are not detectable with an error of $<$0.1 mag at $J$, 
we additionally measured their $J$-band magnitudes with the Stokes $I$ image of $J$-band. 
(c/d) 
Candidates for the exciting star of the UCHII region.  
The positions of the candidates are marked by white circles. 
The position of IRAS 14159-6111 is shown with an orange error ellipse of position. 
North is at the top, east to the left. 
These images have the same field of view. 
(c) 
The three color composite image was constructed from the Spitzer-GLIMPSE 3.5 $\mu$m (blue) image, 
8 $\mu$m (green) image and the Spitzer-MIPSGAL 24 $\mu$m (red) image, 
and 
(d) 
the monochrome image was taken from the Super COSMOS H$_\alpha$ image.
}
\label{fig8}
\end{figure}

The exciting star of the UCHII region is assigned to be a B1 type star from the radio flux, 
although it is unresolved spatially. 
In order to identify the exciting stars, 
we try to search for their near-IR counterparts within 60$\arcsec$ of the IRAS position, 
the UCHII region and the brightest part of the 8 $\micron$ PAH emission. 
In the $H$ versus $H-K_{\rm s}$ diagram (Figure \ref{fig8}a), 
we found five candidates responsible for the UCHII region lying above the reddening line of B2. 
We also made the color-color diagram for these sources (Figure \ref{fig8}b). 
These five sources are also marked  on the three color composite and monochrome images (Figure \ref{fig8}c, d), 
and their properties are presented on Table 1. 
Sources \#4, and \#5 are located outside of the mid-IR nebula, 
while Sources \#1, \#2, and \#3 lie within the brightest part of the nebula. 
Sources \#1, \#2, and \#3 are likely to be embedded B1, B2, and B2 stars, 
that suffer extinction as large as $A_V$ $\sim$40, $\sim$15, and $\sim$40 mag, respectively (Figure \ref{fig8}a, b). 
The position of source \#1 coincides quite well with that of the IRAS source. 
Therefore, 
source \#1 is the most likely candidate responsible for current star formation activities such as the UCHII region, PAH emission, and far- to mid-infrared radiation. 

In our near-IR image (Figure \ref{fig1}b), 
a small stellar aggregation can be seen toward Cloud~B. 
The center of this aggregation is only $\sim$20'' west of the IRAS position, 
and several stars of this aggregation indicate color-excess on the near-IR color-color diagram. 
It is likely that this is a small stellar cluster associated with the exciting star mentioned above \citep[e.g.,][]{testi98}. 

\section{SUMMARY}
We made the near-IR imaging polarimetry toward SFO~74. 
We found that the magnetic field runs along the bright rim in the layer just behind the bright rim, 
and that the whole magnetic field configuration as well as the whole shape 
is roughly symmetric with respect to the symmetric axis of the cloud.
This configuration is most likely due to the UV radiation from the exciting star.
We estimated the magnetic field strength of  $\sim$90 $\mu{\rm G}$ toward 
Region~1 (the tip region just behind the tip bright rim) 
and $\sim$30 $\mu{\rm G}$ toward Region~2 (the inside region on the cloud axis),
implying that the magnetic field strength toward the compressed layer (Region~1) 
have been more enhanced by the UV radiation than Region~2.
The magnetic pressure is roughly comparable to the turbulent pressure,
and thus the magnetic field seems to make a significant contribution to the internal pressure in Region~1.
The mass-to-flux ratio is close to the critical value in Region~1,
indicating the importance of the magnetic field in the dynamics here.
We speculate that the flat-topped shape of SFO~74 is due to the increase of the magnetic field strength just behind the tip.
Since the number of observations that show clear magnetic field structures is still small,
it is very important to 
make more polarimetric observations of BRCs and
to conduct 3-D MHD simulations 
in order to understand the magnetic field effect on the evolution of BRCs. 

\acknowledgments
We are grateful for the support of the staff of SAAO during the observation runs. 
We also thank the anonymous referee for improving the manuscript. 
This work was partly supported by Grants-in-Aid for Scientific Research (24340038 and 24540233) 
from the Ministry of Education, Culture, Sports, Science and Technology of Japan. 
T.K. and K.S. thank Y. Nakajima for assisting in the data reduction with the SIRPOL pipeline package. 
T.K. is Research Fellow of Japan Society for the Promotion of Science. 
M.T. is partly supported by JSPS Grants-in-Aid for Scientific Research. 

\clearpage

\appendix

\section{Comparison of the Polarization Observations in 2013 and 2014}
We observed the northeast region of SFO~74 in 2014 with SIRPOL in addition to the observation in 2013, 
because the total exposure time of 2013 (600 s per wave-plate angle) toward this region was shorter than those (900 s) toward the other regions. 
Therefore, 
we had a chance to examine the 
consistency of the polarization measurements with SIRPOL between two years. 

\begin{figure}
\epsscale{1}
\plotone{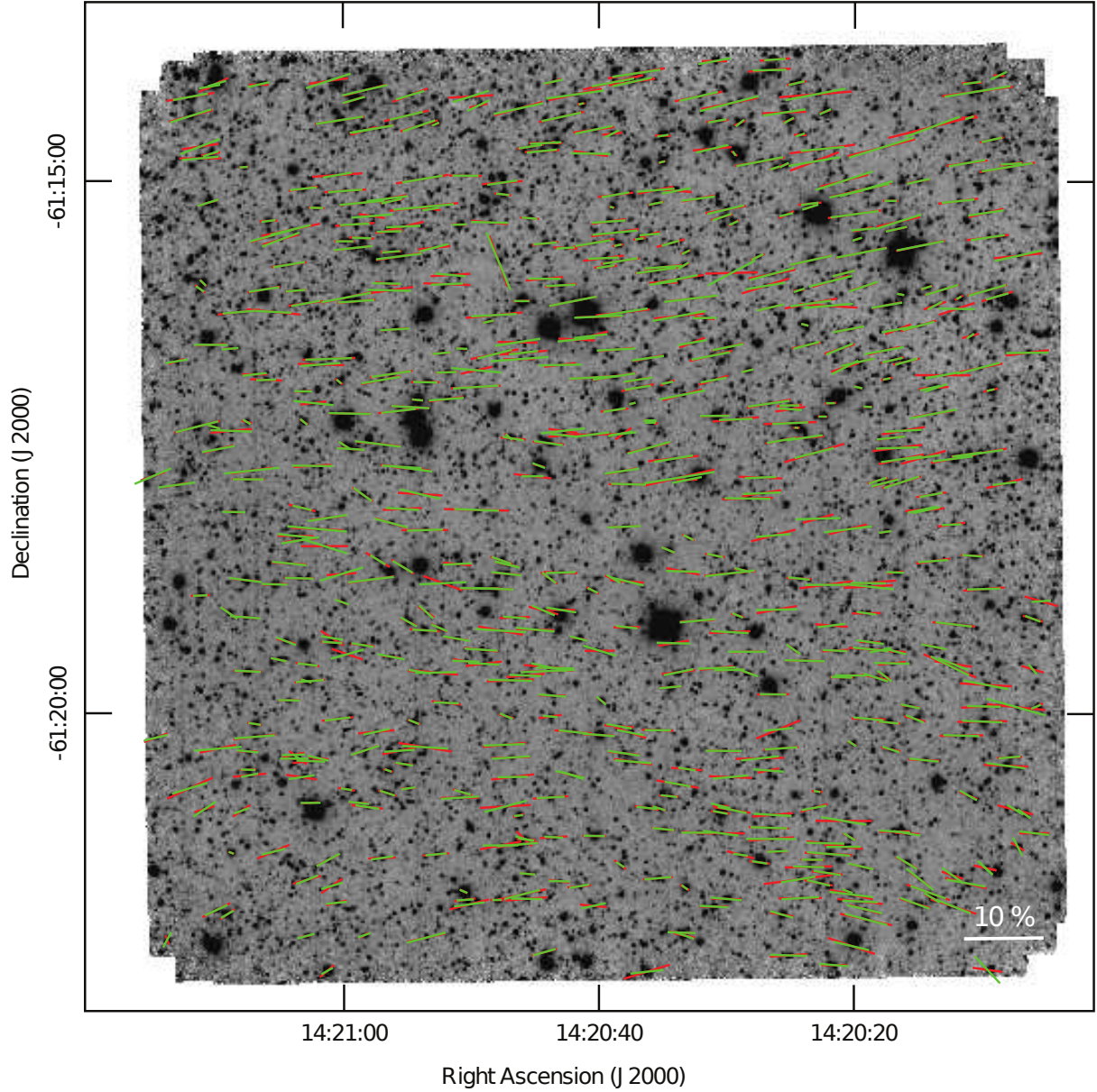} 
\caption{
Polarization vector maps superposed on our $H$-band Stokes $I$ image in 2014. 
Red and green polarization vectors indicate ones obtained in 2013 and 2014, respectively. 
A 10 \% vector is  shown at the right bottom. 
}
\label{vector_map_2013_2014}
\end{figure}

\begin{figure}
\epsscale{0.8}
\plotone{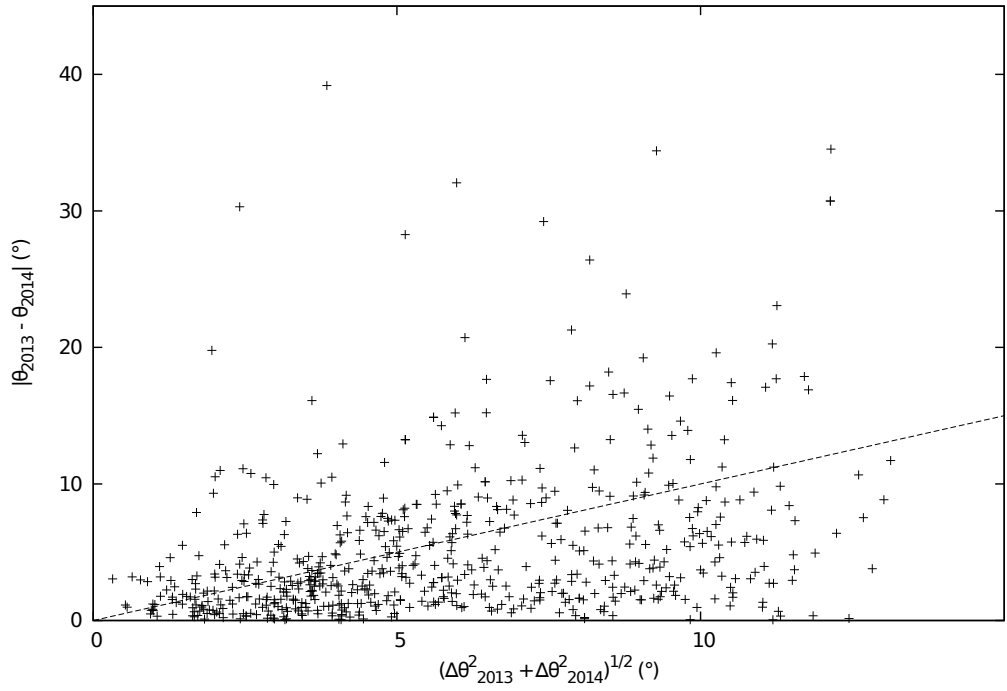}
\caption{
The difference of the polarization angles between 2013 and 2014 versus the root sum square of the polarization angle errors in 2013 and 2014 
for the vectors shown in Figure 14. 
A dashed line shows a line of $|\theta_{2013}-\theta_{2014}|=\sqrt{\Delta\theta^2_{2013}+\Delta\theta^2_{2014}}$. 
}
\label{graph_2013_2014}
\end{figure}

Figure \ref{vector_map_2013_2014} shows the $H$-band vector maps of 2013 and 2014 superposed on the $H$-band $I$ image obtained in 2014. 
We included only the sources that are detected, 
in both 2013 and 2014, with $P/\Delta P>3.0$ ($\Delta \theta\la10\degr$) in the range between the upper and lower limits of Figure \ref{fig3}. 
Red and green vectors indicate those obtained in 2013 and 2014, respectively. 
Most of the green vectors (2014) are likely to be consistent with the red vectors (2013). 
Figure \ref{graph_2013_2014} shows 
the difference of polarization angles for each source between 2013 and 2014, 
where the horizontal axis  is the root sum square of the polarization angle errors of 2013 and 2014 ($ \Delta\theta_{\rm RSS}=\sqrt{\Delta\theta^2_{2013}+\Delta\theta^2_{2014}}$). 
It is found that 
62\% of the sources have the difference of $<\Delta\theta_{\rm RSS}$, 
and that 88\% of them have that of $<2\Delta\theta_{\rm RSS}$. 
As the results of the $J$- and $K_{\rm s}$-bands analysis with the same criteria, 
46\% and 58\% of the sources have the difference of $<\Delta\theta_{\rm RSS}$, respectively. 
These results support the above consistency seen in Figure \ref{vector_map_2013_2014}. 
We also examined the consistency of the polarization degree in $H$-band, 
and found that 60\% of the sources have the difference of the polarization degree of $< \sqrt{\Delta P^2_{\rm 2013}+\Delta P^2_{\rm 2014}}$, 
and that 88\% of them have that of $<2\sqrt{\Delta P^2_{\rm 2013}+\Delta P^2_{\rm 2014}}$.  
For $J$- and $K_{\rm s}$-bands, 
61\% and 55\% of the sources have the difference of $< \sqrt{\Delta P^2_{\rm 2013}+\Delta P^2_{\rm 2014}}$, respectively. 
Therefore, 
we concluded that 
the polarization measurements with SIRPOL are consistent between two years. 

\section{Transformation Angle to the Equatorial Coordinate System}

We had a chance to obtain polarimetric data of the R Mon nebula \citep[e.g.,][]{min91}, 
which is known as the object having centrosymmetic vector pattern due to the scattered light in the outer area of the nebula. 
This object enables us to examine the transformation angle from the SIRPOL coordinate system to the equatorial coordinate system. 
SIRIUS camera is a stationary instrument, 
and SIRPOL is attached to SIRIUS. 
Because the polarimeter of SIRPOL is designed to be attached always the same way with the slide guide settled on the SIRIUS-side flange \citep{kan06}, 
the transformation angle is considered to be highly reproducible. 
With these data, however, 
we tried to make sure whether the transformation angle of $105\degr$ with an accuracy of $\la3\degr$ \citep{kan06} is surely applicable at present. 

With SIRPOL, 
$JHK_{\rm s}$ polarimetric observation of  the R Mon nebula was made on 2014 March 29.  
We obtained 10 dithered, 
each 10 s long, 
at four wave-plate angles as one set of observations and repeated this six times. 
Thus, 
the total on-target exposure time was 600 s per each plate wave-plate angle. 
Sky images were also obtained between target observations 
and were used for medium sky subtraction. 
Twilight flat-field images were obtained at the beginning and/or end of the observations. 
R Mon is located in the peak of the cometary nebula NGC~2261 and is known as the source illuminating the nebula \citep{min91}, 
and we determined the coordinate of R Mon as a peak position from our $J$-band Stokes $I$ image, $(\rm{R.A., Dec.})_{J2000}=(6^{\rm h}39^{\rm m}10\fs0, +8\degr44\arcmin09\farcs7)$. 

\begin{figure}
\epsscale{1}
\plotone{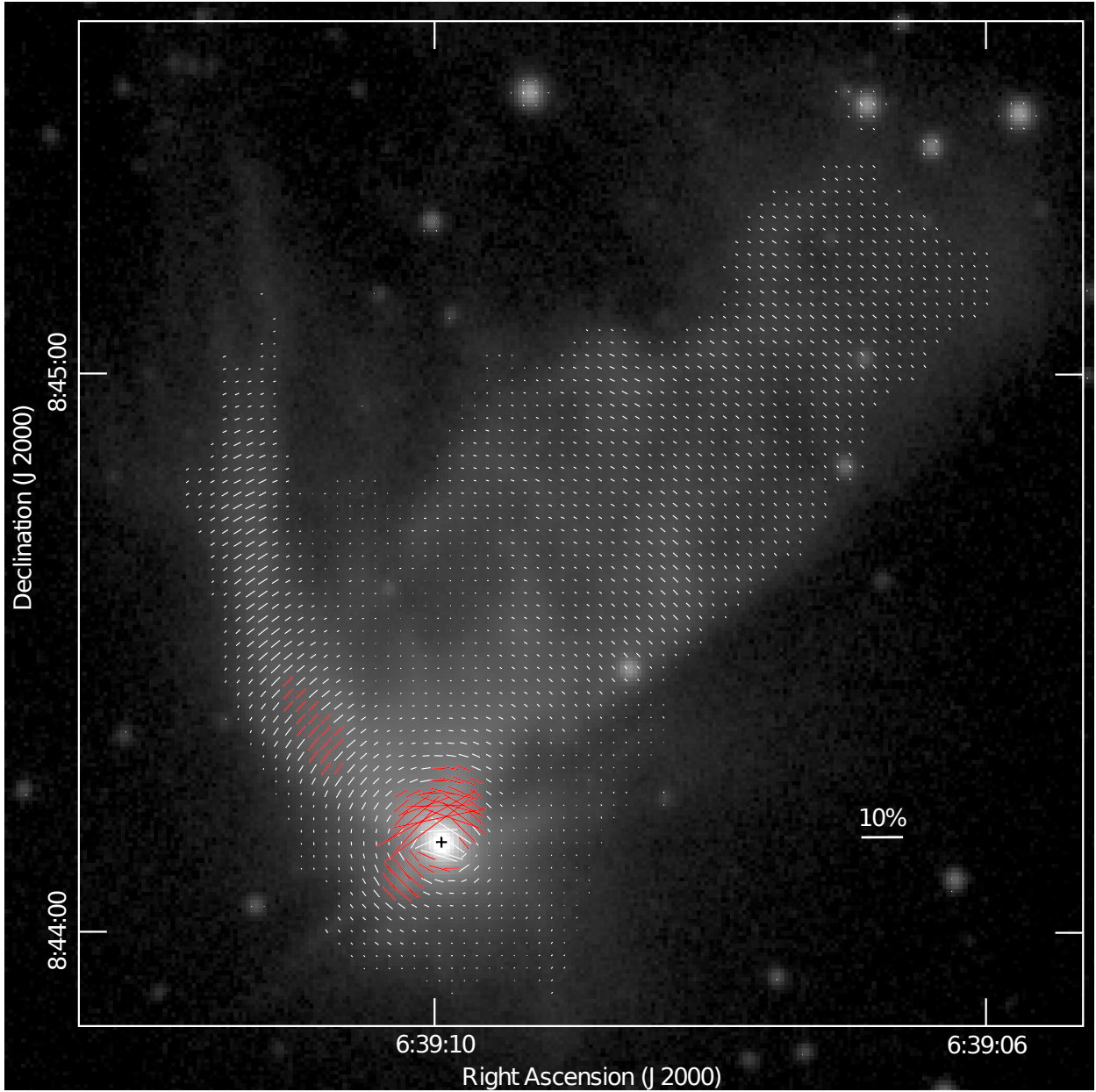}
\caption{
$H$-band polarization vector map of the R~Mon nebula superposed on the $H$-band Stokes $I$ image with a logarithmic scale. 
A 10 \% vector is shown in the lower right corner. 
The vectors were made by $3\times3$ pixel binning. 
A black cross represents the position of R~Mon. 
All vectors are shown in the region where the nebula intensity is greater than $10\sigma$ above the mean sky level.
The vectors with a polarization degree greater than 3 \% are indicated in red, 
and were used to examine the transformation angle from the SIRPOL coordinate system to the equatorial coordinate system, 
excluding the vectors within a radius of $2''$ from the position of R~Mon.
}
\label{R_Mon}
\end{figure}

\begin{figure}
\epsscale{0.8}
\plottwo{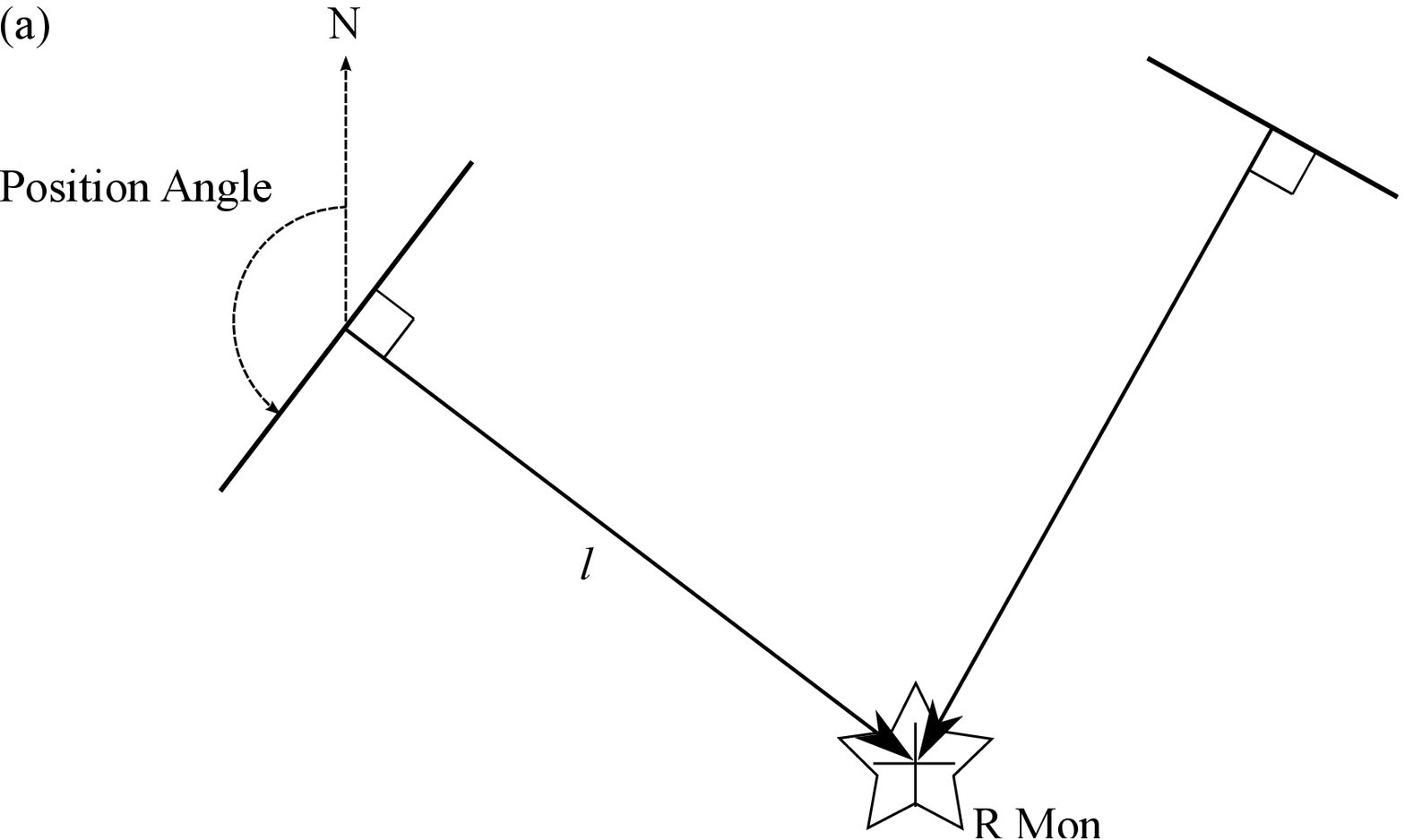}{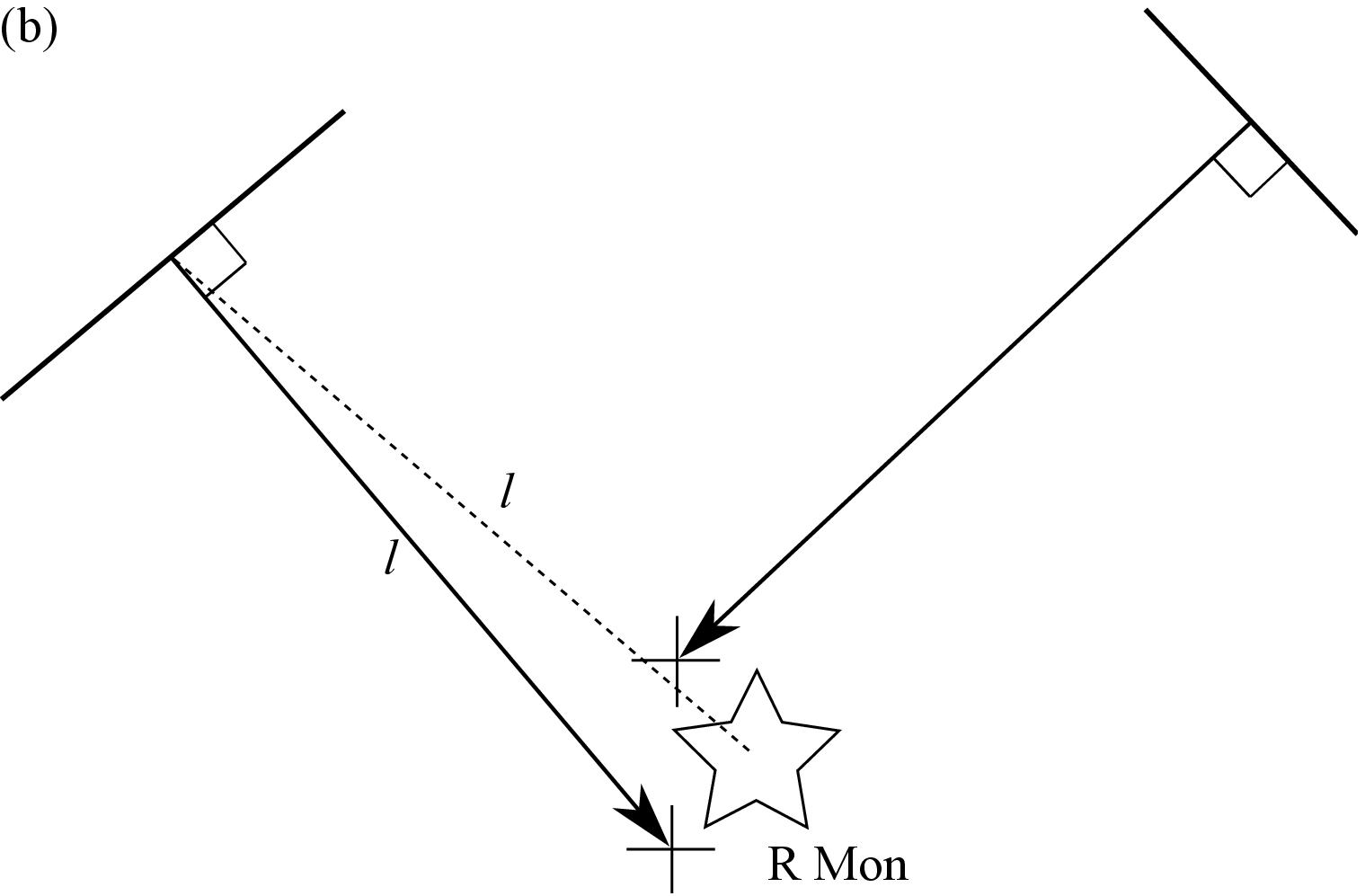}
\caption{
Schematic drawing for a geometrical configuration between a polarization vector and an illuminating source.  
(a) 
In case that a transformation angle from the SIRPOL coordinate system to the equatorial coordinate system is appropriate, 
the terminal point of the normal vector to the polarization vector, with a length $l$ equal to the distance between the illuminating source and the measured point of the polarization angle, 
should coincide with the illuminating source. 
(b) 
If this is not the case,  
the terminal point do not coincide with the illuminating source. 
}
\label{point}
\end{figure}

\begin{figure}
\epsscale{1}
\plotone{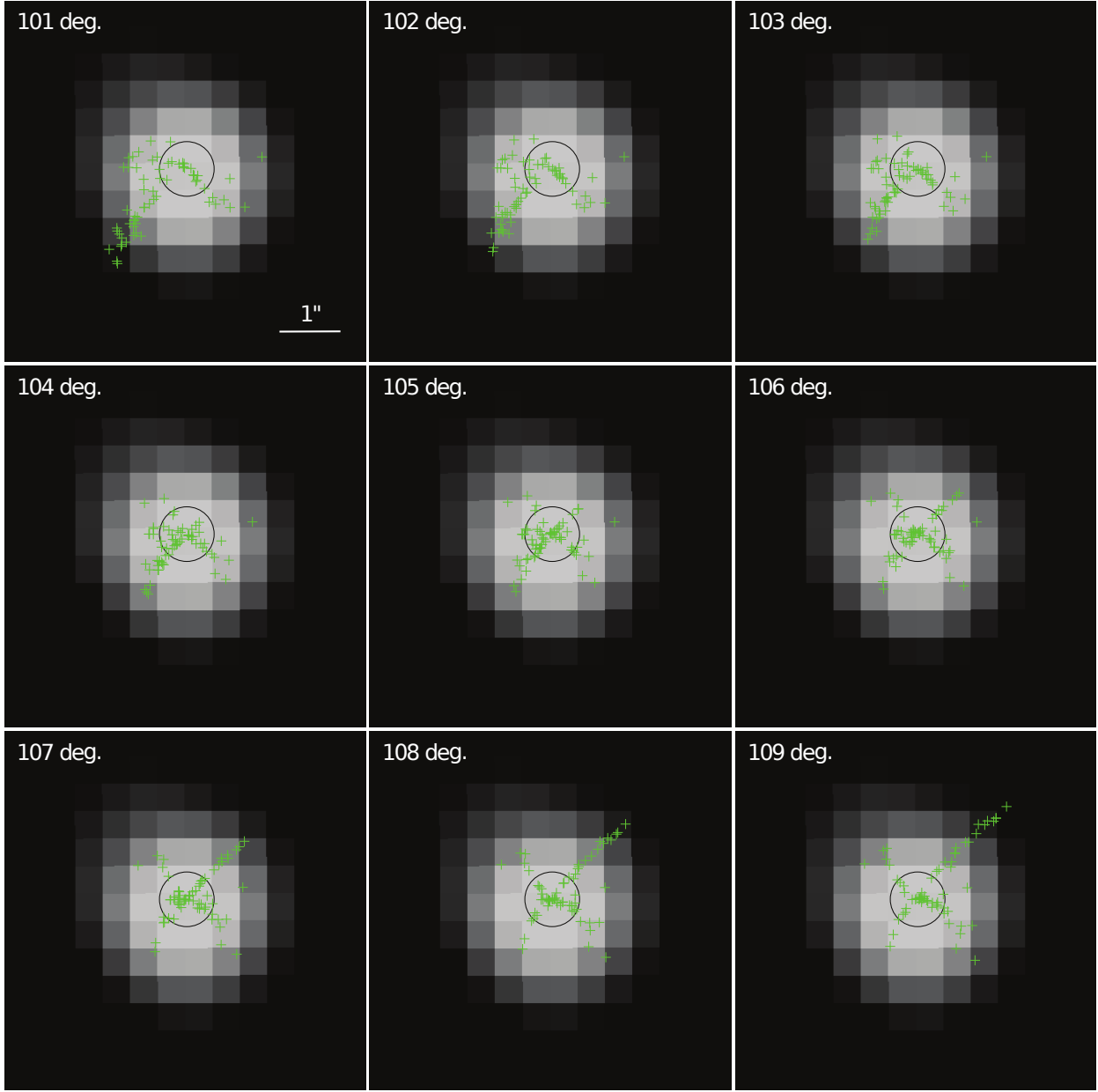}
\caption{ 
Distribution of the terminal points of the normal vectors to the polarization vectors (see the text). 
The center of the black circle shows the position of R Mon. 
On the lower right corner of the 101$\degr$ panel, 
the scale size of 1$\arcsec$ is shown.
}
\label{101-109}
\end{figure}

Figure \ref{R_Mon} shows the $H$-band polarization vector map of R Mon superposed on the $H$-band Stokes $I$ image. 
Here, 
we can see a centrosymmetric vector pattern around R Mon. 
As indicated in Figure \ref{point}, 
if the transformation angle is appropriate, 
the terminal points of the normal vectors to the polarization vectors ideally coincide with the illuminating source. 
In practice, 
however, 
they do not exactly coincide with the illuminating source, 
and are scattered around the illuminating source. 
If the transformation angle is not appropriate, 
scattering of the terminal points should become larger. 
Therefore, 
to determine the most appropriate transformation angle, 
we test  the transformation angles with a range of 101$\degr$-109$\degr$.  
Here, 
we used only the polarization vectors with polarization degrees of $\ga$ 3\% (red vectors in Figure \ref{R_Mon}).  
As indicated in Figure \ref{101-109}, 
the transformation angles of 104$\degr$ - $106\degr$ appear to be more appropriate than those of 101$\degr$-103$\degr$ and 107$\degr$-109$\degr$. 
In $J$-band and $K_{\rm s}$-band analyses that were executed in the same way as the $H$-band analysis, 
the transformation angle of 104$\degr$-106$\degr$ is also likely to be appropriate. 
We conclude that  
the transformation angle of 105$\degr$ reported by \citet{kan06} is valid at present, 
and therefore we adopted 105$\degr$ as the transformation angle of SIRPOL in this paper.

\clearpage

\clearpage

\clearpage

\clearpage

\clearpage

\clearpage

\clearpage

\clearpage

\clearpage

\clearpage

\clearpage

\clearpage

\clearpage

\clearpage

\clearpage

\clearpage

\clearpage

\clearpage

\clearpage

\begin{table}
\scriptsize
\begin{center}
\caption{Candidates for Exciting Stars of Ultracompact HII Region \label{tbl-1}}
\begin{tabular}{cccccccccc}
\tableline\tableline
Near-IR & R.A.(J2000) & Dec.(J2000) & $J$ & $H$ & $K_{\rm s}$ & Optical & Near-IR & Nebula \\
source & & & (mag) & (mag) & (mag) & Counterpart & Sp. Type & Association \\
\tableline
1 & 14$^{\rm h}$19$^{\rm m}$42\fs87 & -61$\degr$25$\arcmin$12$\arcsec$ & 19.25$\pm$0.35 & 14.09$\pm$0.04 & 11.61$\pm$0.02 & No & B1 & Yes \\
2 & 14$^{\rm h}$19$^{\rm m}$40\fs92 & -61$\degr$25$\arcmin$05$\arcsec$ & 12.76$\pm$0.04 & 11.16$\pm$0.04 & 10.28$\pm$0.02 & No  & B2 &Yes? \\
3 & 14$^{\rm h}$19$^{\rm m}$44\fs42 & -61$\degr$25$\arcmin$35$\arcsec$ & 19.07$\pm$0.19 & 14.58$\pm$0.04 & 12.16$\pm$0.02  & No  & B2 & Yes \\
\tableline
4 & 14$^{\rm h}$19$^{\rm m}$48\fs23 & -61$\degr$24$\arcmin$57$\arcsec$ & 12.04$\pm$0.04 & 10.68$\pm$0.04 & 9.96$\pm$0.02 & Yes  & M & foreground \\
5 & 14$^{\rm h}$19$^{\rm m}$36\fs44 & -61$\degr$25$\arcmin$52$\arcsec$ & 14.18$\pm$0.04 & 11.56$\pm$0.04 & 10.31$\pm$0.02 & No & giant & background \\
\tableline
\end{tabular}
\end{center}
\end{table}

\end{document}